\documentclass[prb,twocolumn,superscriptaddress,showpacs,floatfix]{revtex4-1}

\usepackage[final]{graphicx}
\usepackage{amsfonts,amssymb,amsmath}
\usepackage{prettyref}

\usepackage{braket}

\usepackage{color}
\definecolor{red}{rgb}{1.0,0.0,0.0}
\definecolor{blue}{rgb}{0.0,0.0,1.0}
\definecolor{green}{rgb}{0.0,1.0,0.0}

\newcommand{\pref}{\prettyref}

\newrefformat{cha}{Chapter~\ref{#1}}
\newrefformat{sec}{Section~\ref{#1}}
\newrefformat{fig}{Fig.~\ref{#1}}
\newrefformat{eq}{Eq.~(\ref{#1})}
\newrefformat{tab}{Tab.~\ref{#1}}
\newrefformat{app}{Appendix~\ref{#1}}

\newcommand{\mrm}[1]{\ensuremath{\mathrm{#1}}}

\newcommand{\Tr}{\mathrm{Tr}}

\newcommand{\bmm}[1]{\begin{bmatrix}#1\end{bmatrix}}

\newcommand{\hc}[1]{\ensuremath{#1^\dagger}}

\newcommand{\ti}{\tilde}

\newcommand{\ree}{\mathrm{Re}}
\newcommand{\imm}{\mathrm{Im}}

\newcommand{\bs}[1]{\ensuremath{\boldsymbol{#1}}}
\newcommand{\cd}{\cdot}

\newcommand{\al}[1]{\begin{align} #1 \end{align}}

\newcommand{\ml}[1]{\begin{multline} #1 \end{multline}}
\newcommand{\bT}[1]{\ensuremath{\left\{ #1 \right\}}}
\newcommand{\bP}[1]{\ensuremath{\left( #1 \right)}}
\newcommand{\bS}[1]{\ensuremath{\left[ #1 \right]}}

\newcommand{\e}[1]{e^{#1}}
\newcommand{\ee}{\exp}
\newcommand{\abs}[1]{\lvert #1 \rvert}

\newcommand{\pd}[1]{\partial_{#1}}

\newcommand{\bsk}{\boldsymbol{k}}

\newcommand{\hb}{\hbar}
\newcommand{\hbm}{\hbar^{-1}}

\newcommand{\om}{\omega}
\newcommand{\omm}[1]{\omega_{#1}}
\newcommand{\sig}{\sigma}
\newcommand{\sigg}[1]{\sigma_{#1}}

\newcommand{\asig}[1]{\braket{\sigma_{#1}(t)}}

\begin{document}

\title{Microscopic theory of phonon-induced effects on semiconductor quantum dot decay dynamics in cavity QED}

\author{P. Kaer}\email{per.kaer@gmail.com}
\affiliation{DTU Fotonik, Department of Photonics Engineering, Technical University of Denmark, Building 345, 2800 Kgs. Lyngby, Denmark}
\author{T. R. Nielsen}
\affiliation{DTU Fotonik, Department of Photonics Engineering, Technical University of Denmark, Building 345, 2800 Kgs. Lyngby, Denmark}
\author{P. Lodahl}
\affiliation{Niels Bohr Institute, University of Copenhagen, Blegdamsvej 17, 2100 Copenhagen, Denmark}
\author{A.-P. Jauho}
\affiliation{Center for Nanostructured Graphene (CNG), Department of Micro- and Nanotechnology Engineering, Technical University of Denmark, Building 344, 2800 Kgs. Lyngby, Denmark}
\author{J. M{\o}rk}
\affiliation{DTU Fotonik, Department of Photonics Engineering, Technical University of Denmark, Building 345, 2800 Kgs. Lyngby, Denmark}

\date{\today}

\begin{abstract}
We investigate the influence of the electron-phonon interaction on the decay dynamics of a quantum dot
coupled to an optical microcavity.
We show that the electron-phonon interaction has important consequences on the dynamics, especially when the quantum dot and cavity are tuned out of resonance, in which case the phonons may add or remove energy leading to an effective non-resonant coupling between quantum dot and cavity.
The system is investigated using two different theoretical approaches:
(i) a second-order expansion in the bare phonon coupling constant, and (ii) an expansion in a polaron-photon coupling constant, arising from the polaron transformation which allows an accurate description at high temperatures.
In the low temperature regime we find excellent agreement between the two approaches.
An extensive study of the quantum dot decay dynamics is performed, where important parameter dependencies are covered.
We find that in general the electron-phonon interaction gives rise to a greatly increased bandwidth of the
coupling between quantum dot and cavity.
At low temperature an asymmetry in the quantum dot decay rate is observed, leading to a faster decay when the quantum dot has a larger energy than to the cavity.
We explain this as due to the absence of phonon absorption processes.
Furthermore, we derive approximate analytical expressions for the quantum dot decay rate, applicable when the cavity can be adiabatically eliminated.
The expressions lead to a clear interpretation of the physics and emphasizes the important role played by the
effective phonon density, describing the availability of phonons for scattering, in quantum dot decay dynamics.
Based on the analytical expressions we present the parameter regimes where phonon effects are expected to be important.
Also, we include all technical developments in appendices.
\end{abstract}


\maketitle




\section{Introduction}
The study of cavity QED (cQED) has for decades been an important topic in physics. Originally, the main ingredients were atoms, highly confined modes of light, and their mutual interaction. Recent years have seen a rebirth of cQED, but with focus shifted from the pure setting of atoms and cavities, to the complex setting of many-body physics found in semiconductor solid-state systems. A major driving force behind this shift is the advent of quantum information technologies \cite{Knill2001}, with the requirements of applications pushing for the exploration of new material platforms. A scalable all-solid-state platform, where the interaction between light and matter can be engineered and controlled to a high degree \cite{Reinhard2011,Munsch2012,Lermer2012}, could help usher practical devices employing quantum information technologies.

A solid-state platform, however, also poses new challenges owing to its inherent many-body nature, namely the effect of the environment on the fragile quantum states of light and matter and their coherent interaction, which are essential for many applications. Several recent studies \cite{Winger2009,Hohenester2009c,Kaer2010} have shown that simple concepts useful in understanding atomic cQED systems break down on both a quantitative and qualitative level for all-solid-state cQED systems. The two main reasons for the departure from the usual picture are (i) the impossibility of quantum emitters in the solid-state to be described as simple two-level systems and (ii) the stronger coupling to structured environments in the form of, e.g., phonons and electronic inter-particle Coulomb interactions.

For an all-solid-state cQED system consisting of a semiconductor quantum dot (QD) and an optical microcavity especially the interaction with phonons has attracted a considerable amount of attention. It has been shown to influence cQED emission spectra \cite{Milde2008,Winger2009,Calic2011,Hughes2011}, to give rise to detuning dependent spectral asymmetries in QD lifetimes \cite{Hohenester2009c,Kaer2010,Madsen}, as well as yielding unexpected broadening mechanisms in connection with Mollow triplets for coherently driven systems\cite{Majumdar2011,Roy2011a,Ulrich2011,Moelbjerg2012}.

The majority of studies has focused on the effect of phonons in the spectral domain, where typically the spectrum of the emitted light from the entire cQED system is collected and analyzed. However, for quantitative studies, measurements in the temporal domain are in many cases expected to be superior \cite{Madsen2011} due to their  insensitivity towards collection efficiencies. The study of spontaneous emission decay has been employed to probe the environment in which the emitter is emerged into, be it, e.g., electromagnetic \cite{Wang2011} or plasmonic \cite{Andersen2010a} in nature.

In a previous study \cite{Kaer2010} we showed how, at low temperatures, the phonon interaction gives to a significantly faster decay of an excited QD, whose transition frequency is blue-shifted relative to the cavity, as compared to a red-shifted QD. In addition, coupling to phonons gives rise to a  renormalization of the light-matter coupling strength. Similar results have independently been obtained by others \cite{Hohenester2009c,Hohenester2010}.  It was argued that the non-trivial phonon effects could only be explained if the phonons were treated as interacting with the electron-photon quasi-particle, the polariton, and not with the bare electron \cite{Kaer2010}.

Here, we present the details of the theory developed in Ref.~\onlinecite{Kaer2010} and expand the treatment by comparing to an alternative method, more appropriate for higher temperatures. Excellent agreement between the two methods is found in the low-temperature regime, which is of our primary concern. We perform an extensive parameter study, providing, a good picture of the dynamics in different regimes. We furthermore derive an analytical expression for the QD decay rate, which makes the involved physical processes apparent. The analytical expression has very recently been used to experimentally map out the effective phonon density \cite{Madsen}. Furthermore, it inspired to a novel approach for decreasing phonon-induced dephasing in cQED systems \cite{Nysteen}. Furthermore, we provide a simple explanation as to why phonon-induced asymmetries have largely remained unobserved in experimental data until recently.

The paper is organized as follows. In \pref{sec:model_system} we describe the model, emphasizing the interaction with phonons, and introduce the the polaron transformation enabling the treatment of higher temperatures. Sec. \ref{sec:EOM_main} gives a detailed description of the theoretical formalisms employed as well as providing a physical interpretation of the resulting equations of motion. We pursue two methods; The first is based on a second-order expansion in the phonon coupling, yielding simple equations that provide valuable insight into the physics. The second method employs a partially infinite order expansion in the phonon coupling, based on the polaron transformation, which leads to more accurate results but less physical insight. In \pref{sec:results_main} we present a detailed parameter investigation of the models, covering experimentally relevant parameter regimes and discuss the physics of the system. Furthermore, we perform a large detuning expansion and obtain analytical expressions for the total decay rate of the QD, which explicitly accounts for the different contributions to the system decay and make the physical processes very apparent. Finally in \pref{sec:summary_conclusion} we summarize and conclude.

\section{Model system}\label{sec:model_system}
\begin{figure}[h]
 \centering
 \includegraphics[width=0.47\textwidth]{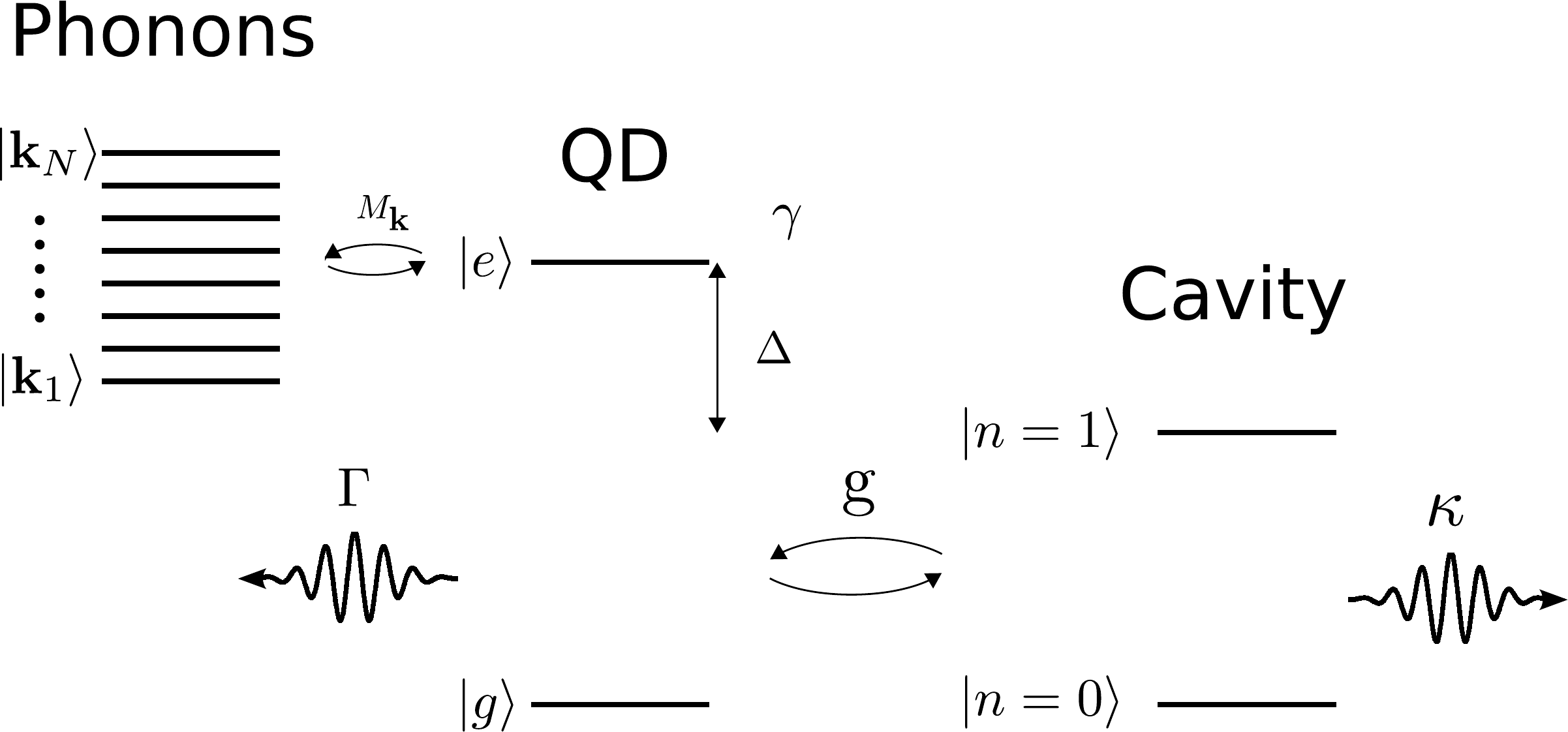}
\caption{Schematic of the cavity QED system including the phonon interaction. The QD-cavity coupling strength is $g$ and the QD-phonon interaction matrix elements are $M_{\bs{k}}$. The rates $\Gamma$ and $\kappa$ yield decay of the QD and cavity, respectively. Pure dephasing of the QD is included through $\gamma$ and $\Delta$ is the QD-cavity detuning.}
\label{fig:QED_system}
\end{figure}

In this section we present the model used to describe the cQED system, including the interaction with phonons. The system is illustrated schematically in \pref{fig:QED_system}. We also devote a section to the polaron transformation.
\subsection{Cavity QED system}\label{sec:cQED_system}
The part of the system consisting of the QD and cavity can be represented by the Hamiltonian
\al{\label{eq:fundamental_Hamil_cQED}
H_\mrm{cQED} = H_\mrm s+H_{\gamma}+H_\kappa + H_\Gamma.
}
Here $H_\mrm s$ describes the QD-cavity system and $H_{\gamma}$, $H_\kappa$, and $H_\Gamma$ describe various interactions with the environment, included as Lindblad loss terms \cite{Carmichael1999,Breuer2002}, to be discussed below. The QD-cavity Hamiltonian reads
\ml{
  H_\mrm s = \sum_{i=\mrm e, \mrm g}\hb\om_i \hc c_i c_i + \hb\om_\mrm{cav} \hc a a +\hb g (\hc a \hc c_\mrm g c_\mrm e + \hc c_\mrm e c_\mrm g a),
}
where the usual rotating wave and dipole approximations have been applied. The energy of the ground (excited) QD state is $\hb\om_\mrm g$ ($\hb\om_\mrm e$) with corresponding fermionic operators $\hc c_\mrm g, c_\mrm g$  ($\hc c_\mrm e, c_\mrm e$), the energy of the cavity photon is $\hb\om_\mrm{cav}$ with corresponding bosonic operators $\hc a, a$, and $g$ is the interaction strength between the cavity photon and the electron in the QD. As we are only concerned with the dynamics of the system on the single photon level, it is advantageous to project the second quantized Hamiltonian, presented above, onto a lower dimensional Hilbert space. An appropriate basis to span this part of the total Hilbert space is the following: $\bT{\ket 1 = \ket{\mrm e, n=0},\ket 2 = \ket{\mrm g, n=1},\ket 3 = \ket{\mrm g, n=0}}$, where $n$ refers to the number of cavity photons. If we project onto this basis and shift to a rotating frame, we can write the QD-cavity system Hamiltonian as
\al{\label{eq:Hs_def_first}
  H_\mrm s = \hb\Delta\sig_{11}+\hb g (\sig_{12}+\sig_{21}),
}
where $\Delta =\om_\mrm{e} -\om_\mrm{g} - \om_\mrm{cav} = \om_\mrm{eg} - \om_\mrm{cav}$ is the QD-cavity detuning and $\sig_{pq} = \ket p \bra q$ is the standard projection operator. The detailed steps are given in \pref{app:fundamental_Hamil}.

The remaining terms in $H_\mrm{cQED}$ all give rise to different forms of losses, which we include through the Lindblad formalism often employed in density matrix theory. The Hamiltonian $H_{\gamma}$ represents pure dephasing processes, with rate $\gamma$, for all transitions connected to the QD, whereas the Hamiltonians $H_\kappa$ and $H_\Gamma$ account for population decay from the cavity and QD by rates $\kappa$ and $\Gamma$, respectively \cite{Carmichael1999}. These rates are taken as parameters with experimentally relevant values.

\subsection{Phonons}
The Hamiltonians involving phonons are given by
\al{\label{eq:H0_ph_def}
H_\mrm{0,ph} &= \sum_{\bs k} \hb\om_\mrm{\bs k}\hc b_\mrm{\bs k}b_\mrm{\bs k},\\
H_\mrm{e-ph} &= \sum_{\bs k} \bP{M_\mrm{gg}^\mrm{\bs k}\hc c_\mrm g c_\mrm g+M_\mrm{ee}^\mrm{\bs k}\hc c_\mrm e c_\mrm e}(\hc b_\mrm{-\bs k}+b_\mrm{\bs k}),
}
where $H_\mrm{0,ph}$ describes the free phonons and $H_\mrm{e-ph}$ describes the electron-phonon interaction. It should be noticed that we assume bulk phonon modes\cite{Krummheuer2002,Roszak2005,Krgel2005,Grosse2007,Milde2008,Stock2011}. The LA phonon dispersion relation is assumed to be linear in the relevant energy range, $\om_{\bs k} = c_\mrm{s}\abs{\bs k}$, with $c_\mrm{s}$ the speed of sound. $\hc b_\mrm{\bs k},b_\mrm{\bs k}$ are the bosonic operators for the phonons. The matrix element $M^{\bs k}_\mrm{\nu\nu}$ in the electron-phonon interaction is\cite{Krummheuer2002,Roszak2005,Krgel2005,Milde2008,Grosse2007}
\al{\label{eq:Mk_full_def}
  M^{\bs k}_\mrm{\nu\nu} = \sqrt{\frac{\hbar k}{2 d c_\mrm s V}}D_\nu \int d\bs r |\phi_\nu(\bs r)|^2 \e{-i\bs k \cd \bs r},
}
where $d$ the is mass density, $c_\mrm s$ is the speed of sound in the material, $V$ is the phonon quantization volume, $D_\nu$ is the deformation potential, and $\phi_\nu(\bs r)$ is the electronic wavefunction for the state involved in the phonon process. We neglect the polar coupling to longitudinal optical (LO) phonons due to their large energies, $\sim 37~\mrm{meV}$, compared to the energies involved in this model, and hence very non-resonant nature. Also, we neglect the piezoelectric coupling to LA phonons, which has been shown to have a small effect for the present system \cite{Krummheuer2002}.

To model the QD wavefunctions, we consider harmonic confinement in the direction perpendicular to the growth direction\cite{Wojs1996} and infinite potentials in the growth direction. This implies wavefunctions for both the ground and excited state of the form
\ml{
\phi_\nu(\bs r) =\frac{2^{1/2}}{\pi^{1/2}l_{xy,\nu}l^{1/2}_\mrm{eff,z}} \exp[-(x^2+y^2)/(2l^2_{xy,\nu})] \\ \times\cos (\pi z/l_\mrm{eff,z}), ~ \abs z \leq l_\mrm{eff,z}/2
}
where the confinement lengths $l_{xy,\nu}$ and $l_\mrm{eff,z}$ can be chosen to model a specific system. We choose QD and phonon parameters suitable for typical InGaAs systems \cite{QD_phonon_parameters}.

If we take advantage of the fact that we only consider a single electron, i.e. $\hc c_\mrm g c_\mrm g+\hc c_\mrm e c_\mrm e = 1$, and project onto the basis introduced above, we obtain
\al{\label{eq:Heph_def_first}
  H_\mrm{e-ph} = \sig_{11}\sum_{\bs k} M^\mrm{\bs k}(\hc b_\mrm{-\bs k}+b_\mrm{\bs k})=\sig_{11}B,
}
where we introduced the effective matrix element
\al{
M^\mrm{\bs k}=M_\mrm{ee}^\mrm{\bs k}-M_\mrm{gg}^\mrm{\bs k}.
}
The details are presented in \pref{app:fundamental_Hamil}.

\subsection{The polaron transformation}\label{sec:pol_trans_main}
We start from the following Hamiltonian\cite{pol_lindblad_rates}
\ml{
H = \hb\Delta\sig_{11} +\hb g (\sig_{12}+\sig_{21}) \\
+ \sig_{11}\sum_{\bs k} M^\mrm{\bs k}(\hc b_\mrm{-\bs k}+b_\mrm{\bs k}) + \sum_{\bs k} \hb\om_\mrm{\bs k}\hc b_\mrm{\bs k}b_\mrm{\bs k},
}
obtained by combining Eqs. (\ref{eq:Hs_def_first}), (\ref{eq:H0_ph_def}), and (\ref{eq:Heph_def_first}).
We then apply the polaron transformation \cite{Wurger1998,Wilson-Rae2002,Brandes2005,Hohenester2010,Roy2011a}, where an operator $O$ transforms as
\al{\label{eq:POL_trans_def}
\bar O = \e{S} O \e{-S}
}
where
\al{
S = \sig_{11}C,\quad C = \sum_{\bs k} \lambda_{\bs k}(\hc b_\mrm{-\bs k}-b_\mrm{\bs k}),\quad \lambda_{\bs{k}} = \frac{M^\mrm{\bs k}}{\hbar \om_{\bs k}}.
}
The idea behind the transformation is to remove the term linear in the phonon operators in order to arrive at a set of equations that is easier to treat. Physically, the transformation shifts the phonon modes according to the presence of the electron, determined by the operator $\sig_{11}$. From the exponential nature of the transformation operator $\e S$, phonon processes are included to infinite order. This has the consequence that multi-phonon effects are easily included in the theory, allowing for the description of experiments performed at high temperatures. We use the bar to signify the transformed frame. The Hamiltonian in the polaron frame becomes
\al{
\bar H = \bar H_{\mrm s'} + \bar H_{\mrm s'-\mrm{ph}'} + H_{\mrm{0,ph}},
}
with
\begin{subequations}\label{eq:Hpol_individual}
\al{
\bar H_{\mrm s'} &= \hb\Delta\sig_{11}+\hb g \braket X(\sig_{12}+\sig_{21}), \\
\bar H_{\mrm s'-\mrm{ph}'} &= \hb g (\sig_{12}\delta X_++\sig_{21}\delta X_-), \\
H_{\mrm{0,ph}} &= \sum_{\bs k} \hb\om_\mrm{\bs k}\hc b_\mrm{\bs k}b_\mrm{\bs k}.
}
\end{subequations}
It should be noted that a constant energy shift, induced by the phonons, has been absorbed in the QD-cavity detuning $\Delta$, see \pref{eq:polaron_shift}. Also, new phonon related operators have been introduced
\al{
\delta X_\pm &= X_\pm-\braket{X},\\
X_\pm &= \e{\pm C},
}
where it holds that
\al{
\braket X &= \braket{X_\pm}.
}
The brackets denote the expectation value with respect to the thermal density matrix for the phonons, more precisely $\braket{\cdots}=\Tr_\mrm{ph}\bT{\rho_{\mrm{ph},0} \cdots}$. The detailed derivation can be found in \pref{app:polaron_trans_hamil} and various relevant properties of the operators $X_\pm$ are described in \pref{app:prop_phon_op}. Due to the polaron transformation, the division of the total Hamiltonian into a QD-cavity system part and a phonon part is no longer possible. Indeed, the new system Hamiltonian, $\bar H_{\mrm s'}$, contains the phonon quantity $\braket X$ which is seen to renormalize the light-matter coupling strength $g$. It should also be noted that defining the new system Hamiltonian in this way, we include photon processes to infinite order and respect the detailed balance condition \cite{Carmichael1973}. From the expression for $\braket X$, see \pref{eq:X_avg_def}, it is obvious that $0 < \braket X \leq 1$. The interaction with phonons will thus always decrease the effective light-matter coupling strength as a consequence of this. The new interaction Hamiltonian, $\bar H_{\mrm s'-\mrm{ph}'}$, contains the phonon fluctuation operators $\delta X_\pm$, describing fluctuations of the phonon bath around its equilibrium value, as well as the light-matter coupling strength $g$.




\section{Equations of motion}\label{sec:EOM_main}
In this section we present the theoretical formalism employed for analyzing the system, described by the Hamiltonians of the previous section. The explicit form of the equations of motion is also presented.

\subsection{Time-convolutionless approach}
Our basic approach is to set up an equation of motion for the reduced density matrix (RDM) of the QD-cavity system, where the phonon degrees of freedom are traced out. This is a standard technique \cite{Breuer1999,Carmichael1999,Breuer2002} in which the effect of the reservoir enters through various scattering terms in the equation of motion (EOM) for the RDM. These scattering terms can be derived by two different approaches. In the first, known as the Nakajima-Zwanzig projection operator technique \cite{Breuer2002}, the resulting EOMs have memory: the present state of the system thus depends on the past history. In the second, known as the time-convolutionless approach \cite{Breuer2002} (TCL), the EOMs are time-local and therefore do not have memory, however, the scattering rates become time-dependent. Both of these approaches yield, without further approximations, a non-Markovian description of the dynamics.

In this paper we employ the TCL up to second order in the perturbation, for the following two reasons: The first and most important is that in the limit where the light-matter coupling tends to zero, our model reduces to the so-called independent boson model \cite{Mahan1993}. This model is known to be exactly solvable using a number of methods, one being the second order TCL \cite{Weber2008}. Even though the present model can not be solved exactly using the second order TCL, we expect the result to be more accurate compared to that obtained using the method involving memory integrals, since that method does not lead to the exact solution to second order for $g \rightarrow 0$. Other studies have also shown the TCL to be superior to the corresponding equation with memory \cite{Breuer1999}. The second reason is purely practical, in that time-local equations are simpler to solve than equations containing memory integrals

The EOMs arising from the TCL may be derived in a completely general framework \cite{Breuer2002}, however we follow a less rigorous approach in deriving the TCL and present the resulting formulas in \pref{app:EOM_RDM}.

\subsection{Phenomenological losses}\label{sec:lindblad_losses}
As mentioned in \pref{sec:cQED_system}, we also include interactions with other reservoirs than phonons to simulate a real system with losses. These are included using the Lindblad formalism \cite{Carmichael1999}, where terms of the form
\ml{\label{eq:Lindblad_term}
L\bT{O,\gamma}\rho(t) =\\ -\frac \gamma 2 \bS{\hc O O \rho(t)+\rho(t) \hc O O -2 O \rho(t) \hc O},
}
are added to the EOM for $\rho(t)$, where $\rho(t) = \mrm{Tr}_{\mrm{ph}}\bT{\chi(t)}$ is the RDM for the QD-cavity system, $\chi(t)$ is the density matrix for the total system, and $\mrm{Tr}_{\mrm{ph}}\bT{\cdots}$ denotes the trace operation with respect to the phonon degrees of freedom. The above leads to decay with rate $\gamma$ of the transition corresponding to the operator $O$. This expression may be obtained by taking the white noise, or equivalently zero memory, and zero temperature limit of the scattering terms presented in \pref{app:EOM_RDM}.

The decay of the cavity field through leaky modes is modeled by including the Lindblad term $L\bT{\sigg{32},\kappa}\rho(t)$, the decay of the excited QD through radiative and non-radiative processes is modeled by including $L\bT{\sigg{31},\Gamma}\rho(t)$, and finally a Markovian pure dephasing rate is also included through $L\bT{\sigg{11},2 \gamma}\rho(t)$. We refer to \pref{sec:cQED_system} for notational remarks. Since LA phonons have been included explicitly, and already give rise to a pure dephasing rate, it might seem redundant to introduce an additional pure dephasing channel. However, previous work has demonstrated that excited states for electrons and holes contribute to pure dephasing processes near the ground state transition energy, due to both LA \cite{Muljarov2004,Grange2009a} and LO \cite{Muljarov2007} phonon interactions. Also, including a finite lifetime of either LO and LA phonons, arising, e.g., from anharmonic effects \cite{Jacak2002}, induces a contribution to the pure dephasing rate \cite{Zimmermann2002}. For simplicity, we assume $\gamma$ to be an independent parameter.

\subsection{Notational remarks}
The resulting EOMs we arrive at are all linear in the elements of the RDM. This fact makes it advantageous to formulate the EOMs in the language of linear algebra. This can be achieved by mapping the RDM onto a vector form as follows
\ml{
\braket{\bs{\sigma}(t)} = \bS{\asig{11},\asig{22},\asig{12},\asig{21},\right.\\ \label{eq:u_def}
\left.\asig{23},\asig{32},\asig{13},\asig{31}}^T.
}
Here, $\asig{qp} = \Tr_{\mrm s}\bT{\rho(t)\sigg{qp}}=\rho_{pq}(t)$, where $\Tr_{\mrm s}\bT{\cdots}$ denotes the trace with respect to the QD-cavity basis. The QD ground state population, i.e. $\asig{33}$, has been omitted as it does not matter for the dynamics considered and may be trivially obtained using the  conservation of population. The matrix describing the coupling between different elements can be divided into three main contributions
\al{\notag
\pd t \braket{\bs{\sigma}(t)} &= [M_\mrm{coh}+M_\mrm{Lindblad}+M_\mrm{LA}(t)]\braket{\bs{\sigma}(t)}\\
&=M(t)\braket{\bs{\sigma}(t)},\label{eq:EOM_u_general}
}
where $M_\mrm{coh}$ describes terms originating from the coherent unitary evolution provided by the QD-cavity Hamiltonian, $M_\mrm{Lindblad}$ describes terms from the Lindblad operators, and $M_\mrm{LA}(t)$ describes the time-dependent scattering terms induced by the coupling to LA phonons.

As will be shown, $M(t)$ can be written as two decoupled sub-matrices
\al{
M(t)=
\bmm{
m^{(11)}(t) & 0 \\
0 & m^{(22)}(t)
},
}
where $m^{(11)}(t)$ couples the first four elements of $\braket{\bs\sigma(t)}$, $m^{(22)}(t)$ couples the last four, and all other elements are zero.

In the following two sections we will derive the EOMs for the system using the TCL. We present the equations arising from the Hamiltonian without the polaron transformation, denoted the original frame, and with the polaron transformation, denoted the polaron frame. Employing the polaron transformed Hamiltonian is expected to yield more accurate results compared to the original Hamiltonian, especially for elevated temperatures. However, the equations resulting from the polaron transformation are also more complicated and due to the change of basis harder to interpret physically. On the other hand, the equations arising in the original frame are simple and can be used to gain insight into the physics.
\subsection{Original frame}\label{sec:theory_orig_frame}
In the original frame, i.e. not employing, the polaron transformation, the total Hamiltonian without the Lindblad contributions is
\al{
H = H_\mrm{s}+H_\mrm{0,ph}+H_\mrm{e-ph},
}
where the individual contributions can be found in Eqs. (\ref{eq:Hs_def_first}), (\ref{eq:H0_ph_def}), and (\ref{eq:Heph_def_first}), respectively.
We consider $H_\mrm{e-ph}$ as the interaction Hamiltonian, for which the perturbation expansion is performed. With this choice only the electron-phonon interaction is treated approximately, which is expected to be a good approximation, whereas the electron-photon interaction is treated exactly and the theory is not limited to small values of the light-matter coupling strength $g$.

To write up the TCL EOM for the RDM we use \pref{eq:reduced_DM_eom} and the time-local scattering term given in \pref{eq:scatt_timelocal} and finally add the Lindblad terms discussed in \ref{sec:lindblad_losses} to get \cite{Kaer2010}
\ml{\label{eq:RDM_EOM_orig}
\pd t \rho(t) =-i\hbm\bS{H_\mrm s,\rho(t)}+S_\mrm{LA}(t)\\
+\bP{L\bT{\sigg{32},\kappa}+L\bT{\sigg{31},\Gamma}+L\bT{\sigg{11},2 \gamma}}\rho(t).
}
Written in terms of the operator expectation values $\asig{nm}$, the populations in the QD-cavity system are obtained as follows: The cavity population is $\braket{\hc a(t)a(t) }=\asig{22}$ and the excited QD population $\braket{\hc c_\mrm e(t)c_\mrm e(t) }=\asig{11}$. The off-diagonal elements correspond to different polarizations or coherences in the QD-cavity system, with the relevant one for one-time dynamics being the so-called photon-assisted polarization $\asig{12}$. Remapping the RDM to vector form, we get the following coupling matrices. The coherent terms are
\al{\label{eq:m_coh_orig_11}
m_\mrm{coh}^{(11)}=
\bmm{
0 & 0 & -ig & ig \\
0 & 0 & ig & -ig \\
-ig & ig & i\Delta & 0 \\
ig & -ig & 0 & -i\Delta
},
}
and
\al{\label{eq:m_coh_orig_22}
m_\mrm{coh}^{(22)}=
\bmm{
0 & 0 & ig & 0 \\
0 & 0 & 0 & -ig \\
ig & 0 & i\Delta & 0 \\
0 & -ig & 0 & -i\Delta
},
}
and the Lindblad contributions take the form
\ml{\label{eq:m_lindblad_orig}
\mrm{diag}\bT{M_\mrm{Lindblad}}=\\
-\frac 12\bS{2\Gamma,2\kappa,\Gamma+\kappa+2\gamma,\Gamma+\kappa+2\gamma,\right.\\
 \left.\kappa,\kappa,\Gamma+2\gamma,\Gamma+2\gamma},
}
and finally the phonon induced terms are
\begin{widetext}
\al{\label{eq:s12_ph}
m_\mrm{LA}^{(11)}(t)=
\bmm{
0 & 0 & 0 & 0 \\
0 & 0 & 0 & 0 \\
-i\mathcal{G}^>(t) & i\mathcal{G}^<(t) & -[\gamma_{12}(t)-i\Delta_\mrm{pol}] & 0 \\
i[\mathcal{G}^>(t)]^* & -i[\mathcal{G}^<(t)]^* & 0 & -[\gamma^*_{12}(t)+i\Delta_\mrm{pol}]
},
}
\al{
m_\mrm{LA}^{(22)}(t)=
\bmm{
0 & 0 & 0 & 0 \\
0 & 0 & 0 & 0 \\
i\mathcal{G}^<(t) & 0 & -[\gamma_{13}(t)-i\Delta_\mrm{pol}] & 0 \\
0 & -i[\mathcal{G}^<(t)]^* & 0 & -[\gamma^*_{13}(t)+i\Delta_\mrm{pol}]
}.
}
\end{widetext}
The elements of $m_\mrm{LA}$ will be defined below, Eqs. (\ref{eq:Ggl_def}), (\ref{eq:gam12_def}), and (\ref{eq:gam13_def}), but first we provide a brief discussion of the elements. If one disregards the phonon induced scattering terms, these equations constitute the standard lossy Jaynes-Cummings model including pure dephasing, which has been studied intensely in recent years \cite{Naesby2008,Yamaguchi2008,Auffeves2009,Laucht2009,Gonzalez-Tudela2010}. Let us start by discussing the terms in $m_\mrm{LA}^{(11)}(t)$ in more detail, i.e., the quantities $\gamma_{12}(t)$ and $\mathcal{G}^\gtrless(t)$. If we compare the structure of the phonon scattering term, \pref{eq:s12_ph}, with the non-phonon related terms in the coherent and Lindblad contributions to $M$, a physical interpretation of the effects of phonons becomes apparent.

The rate $\gamma_{12}(t)$ multiplies the photon-assisted polarization and therefore the real part of $\gamma_{12}(t)$ represents pure dephasing of this specific polarization, whereas the imaginary part corresponds to an energy shift. The long-time limit of this energy shift has been subtracted in the form of the quantity $\Delta_\mrm{pol} = \mrm{Im}\bT{\gamma_{12}(\infty)}$, usually referred to as the polaron shift, to provide a consistent expansion in the electron-phonon interaction \cite{counterterm_ref}. This adjustment has been performed everywhere the detuning, $\Delta$, enters and results in an effective QD-cavity detuning close to zero, $\Delta\approx 0$.

The quantities $\mathcal{G}^\gtrless(t)$ multiply the populations of the excited QD-cavity system in such a way that the real part of $\mathcal{G}^\gtrless(t)$ renormalizes the bare light-matter coupling strength $g$. However, in general $\mrm{Re}\bS{\mathcal{G}^>(t)}\neq \mrm{Re}\bS{\mathcal{G}^<(t)}$ and hence the renormalization does not correspond to an overall change in the value of $g$ in the EOM for $\asig{12}$. The imaginary part of $\mathcal{G}^\gtrless(t)$ gives rise to an additional decay or growth of the polarization, depending on the sign of $\mrm{Im}\bS{\mathcal{G}^\gtrless(t)}$, if state 1 or 2 is populated. The influence of the degree of excitation in the QD-cavity system makes this dephasing channel of a different nature than the pure dephasing normally induced by phonons, which is well understood, see e.g. Ref.~\onlinecite{Krummheuer2002}.

From the scattering term \pref{eq:scatt_timelocal} we get
\al{\label{eq:Ggl_def}
\mathcal{G}^\gtrless(t)&=i\hb^{-2}\int_0^t dt'U^*_{11}(t')U_{21}(t')D^\gtrless(t'),\\ \notag
\gamma_{12}(t) &= \hb^{-2}\int_0^t dt'[\abs{U_{11}(t')}^2D^<(t')-\abs{U_{21}(t')}^2D^>(t')]\\ \label{eq:gam12_def}
&= \hb^{-2}\int_0^t dt'[D^<(t')-\abs{U_{21}(t')}^22\ree[D^<(t')]],\\
\Delta_\mrm{pol} &= \mrm{Im}\bT{\gamma_{12}(\infty)},\\ \notag
\gamma_{13}(t) &= \hb^{-2}\int_0^t dt'\abs{U_{11}(t')}^2D^<(t')\\ \label{eq:gam13_def}
 &= \hb^{-2}\int_0^t dt'[D^<(t')-\abs{U_{21}(t')}^2D^<(t')]
}
where it has been used that both $D^\gtrless(t-t')$ and $U_{nm}(t-t')$ only depend on the difference between the two time arguments and further the initial time has been assumed to be zero. The phonon bath correlation functions entering above are defined as
\al{\label{eq:Dgtrless_def}
D^\gtrless(t)&=\sum_{\bs k} \abs{M^{\bs k}}^2 \bS{n_{\bs k}\e{\pm i \omm{\bs k}t}+\bP{n_{\bs k}+1}\e{\mp i \omm{\bs k}t}}\\
&=\sum_{\bs k} \abs{M^{\bs k}}^2 \bS{(2n_{\bs k}+1)\cos (\omm{\bs k}t) \mp i \sin (\omm{\bs k}t)},
}
which are related to the phonon bath operators $B$ in the following way
\al{
D^\gtrless(t-t') = \braket{\ti B(\pm[t-t'])\ti B(0)},
}
and $n_{\bs k}$ is the thermal occupation factor for the $\bsk$'th phonon mode, defined in \pref{eq:avg_phon_nk}. The matrix $U(t)$ is the time-evolution operator for the QD-cavity system which, due to the time-independence of $H_\mrm{s}$, see \pref{eq:Hs_def_first}, can be given as a closed form expression
\al{
U(t)=\exp (-iH_\mrm{s} t/\hb).
}
The products of the elements of $U(t)$ occurring in Eqs. (\ref{eq:Ggl_def}) and (\ref{eq:gam12_def}) can be interpreted as propagators of the QD-cavity system governed by $H_\mrm{s}$, representing the pure lossless Jaynes-Cummings model. This is easily realized by writing the time-evolution of the density matrix for the pure Jaynes-Cummings model as
\al{
\rho^\mrm{JC}(t) = U(t)\rho^\mrm{JC}(0)\hc U(t).
}
If we assume that $\rho^\mrm{JC}(0) = \sigg{kk}$, i.e. the time-evolution starts with the excitation in a single state, we get
\al{
\rho_{nm}^\mrm{JC}(t,\sigg{kk}) = U_{nk}(t)\hc U_{km}(t)=U_{nk}(t)U^*_{mk}(t).
}
The time-evolution of $\rho_{nm}^\mrm{JC}(t,\sigg{kk})$ contains the light-matter coupling, and so do the Jaynes-Cummings propagators entering the phonon induced scattering terms. This leads to the interpretation that the phonons interact not with the bare electron, but rather with an electron-photon quasi-particle \cite{Kaer2010} often referred to as a polariton. Indeed, if we approximate the $U(t)$ matrix in the phonon induced scattering terms with the time-evolution operator obtained for $g=0$, i.e., the non-interacting QD-cavity system, then $U(t)$ becomes strictly diagonal \cite{Harouni2009}. As a consequence $\mathcal{G}^\gtrless(t) = 0$ and $\gamma_{12}(t) = \hb^{-2}\int_0^t dt'D^<(t')$ and the phonon induced scattering terms would not depend on the properties of the QD-cavity system.

\subsection{Polaron frame}
In the RDM formalism we derive an EOM for
\al{
\rho(t) = \Tr_\mrm{ph}\bT{\chi (t)},
}
which is useful for calculating expectation values provided that the operator of interest belongs to the system part of the Hilbert space. In this case we may perform the following operation
\al{
\braket{O(t)} &= \Tr_\mrm{s+ph}\bT{\chi (t)O}=\Tr_\mrm{s}\bT{\Tr_\mrm{ph}\bT{\chi (t)}O}\\
&=\Tr_\mrm{s}\bT{\rho (t)O}.
}
If we now perform an arbitrary basis change operation given by the unitary operator $T$, where $\hc T T= T^{-1}T=I$, the expectation value of the operator $O$ must of course not change, hence
\al{
\braket{O(t)} &= \Tr_\mrm{s+ph}\bT{\chi (t)O}\\
&= \Tr_\mrm{s+ph}\bT{T\hc T\chi (t)T\hc T O T \hc T}\\
&=\Tr_\mrm{s+ph}\bT{\bar\chi (t)\bar O},
}
where the bar signifies the operator in the new basis. In the new basis we may also define a RDM for the system as follows
\al{
\bar \rho(t) = \Tr_\mrm{ph}\bT{\bar \chi (t)}.
}
However, in order for this object to be useful for calculating physical expectation values, we need to be able to perform the following operation
\al{
\braket{O(t)} &= \Tr_\mrm{s+ph}\bT{\bar\chi (t)\bar O}=\Tr_\mrm{s}\bT{\Tr_\mrm{ph}\bT{\bar\chi (t)}\bar O}\\
&=\Tr_\mrm{s}\bT{\bar\rho (t)\bar O}.
}
That is, the basis change should not entangle the system operator with the reservoir degrees of freedom or more formally $\bar O = \bar o_\mrm s \otimes I_\mrm{ph}$, $I_\mrm{ph}$ being the identity operator in the phonon Hilbert space.

In the case of the polaron transformation, see \pref{eq:POL_trans_def}, all system projection operators are left invariant under the polaron transformation, i.e. $\bar\sigma_{nm}=\sigg{nm}$, except for the off-diagonal operators: $\sigg{12}$, $\sigg{13}$, and their hermitian conjugates. This has the consequence, e.g., that the bare electron polarization $\braket{\hc c_\mrm e(t)c_\mrm g(t)}=\Tr_\mrm s\bS{\rho(t)\sigg{13}}$, often used to calculate the linear optical susceptibility, cannot be determined directly within the polaron frame \cite{Wilson-Rae2002}. Fortunately, all  operators needed for our purposes are left invariant.

As the polaron transformed Hamiltonian derived in \pref{sec:pol_trans_main} is expressed in terms of bare QD-cavity operators, the elements of the RDM that are projected out are with respect to the bare QD-cavity system operators and hence do not always correspond to the actual physical elements. To distinguish between expectation values calculated in the polaron and original frame, we introduce the following notation for the expectation values in the polaron frame
\al{
\braket{O(t)}_\mrm p =\Tr_\mrm{s}\bT{\bar\rho (t) O},
}
and as a consequence we get a new vector representation of the RDM in the polaron frame
\ml{
\braket{\bs{\sigma}(t)}_\mrm p = \bS{\asig{11},\asig{22},\asig{12}_\mrm p,\asig{21}_\mrm p,\right.\\
\left.\asig{23},\asig{32},\asig{13}_\mrm p,\asig{31}_\mrm p}^T.
}

The polaron transformed Hamiltonian is given by
\al{
\bar H = \bar H_{\mrm s'} + \bar H_{\mrm s'-\mrm{ph}'} + H_{\mrm{0,ph}},
}
where the individual terms are defined in \pref{eq:Hpol_individual}.
As in the previous section we set up the EOM for the RDM
\ml{\label{eq:RDM_EOM_POL}
\pd t \bar\rho(t) =-i\hbm\bS{\bar H_\mrm{s'},\bar\rho(t)}+\bar S_\mrm{LA}(t)\\
+\bP{L\bT{\sigg{32},\kappa}+L\bT{\sigg{31},\Gamma}+L\bT{\sigg{11},2 \gamma}}\bar\rho(t),
}
where the LA scattering term in this case contains the interaction Hamiltonian $\bar H_{\mrm s'-\mrm{ph}'}$. The coupling matrices in the polaron frame for the coherent and Lindblad terms are identical to those in the original frame, see Eqs. (\ref{eq:m_coh_orig_11}), (\ref{eq:m_coh_orig_22}), and (\ref{eq:m_lindblad_orig}), except that the replacement $g\rightarrow \braket X g$ should be performed in the coherent terms.
The terms arising from the coupling to the LA phonons are
\al{\label{eq:POL_scatt_terms}
m_\mrm{LA}^{(11)}(t)=
\bmm{
-\Gamma_1(t) & +\Gamma_2(t) & -iG_2^*(t) & +iG_2(t) \\
+\Gamma_1(t) & -\Gamma_2(t) & +iG_2^*(t) & -iG_2(t) \\
+iG_1(t) & +iG_1(t) & -\gamma_1(t) & -iG^*_3(t) \\
-iG^*_1(t) & -iG^*_1(t) & +iG_3(t) & -\gamma^*_1(t)
},
}
and
\al{
m_\mrm{LA}^{(22)}(t)=
\bmm{
-\gamma_2(t) & 0 & iG_5(t) & 0 \\
0 & -\gamma^*_2(t) & 0 & -iG^*_5(t) \\
iG_4(t) & 0 & -\gamma_3(t) & 0 \\
0 & -iG^*_4(t) & 0 & -\gamma^*_3(t)
}.
}
All elements are explicitly defined in \pref{app:EOM_polaron}. As these expressions are given in the polaron frame, we can not interpret the different terms as easily as in the original frame. However, we will still note a few differences and similarities. We now see a direct phonon induced lifetime renormalization of states 1 and 2 through $\Gamma_1$ and $\Gamma_2$, as well as several quantities playing a role similar to $\mathcal{G}^\gtrless(t)$ in the original frame, via the $G_n(t)$'s. Also, all polarizations now have a phonon induced pure dephasing rate, given by the quantities $\gamma_n(t)$, associated with them. All quantities are composed from terms of the form
\al{\label{eq:Knmkl_def_main_paper}
K_{nmkl}^\pm(t) = g^2\int_0^t dt' \bar U_{n,m}(t') \bar U^*_{k,l}(t') B_\pm(t'),
}
where
\al{
\bar U(t) = \exp (-i\bar H_\mrm{s'}t/\hbar),
}
is the time-evolution operator with respect to $\bar H_\mrm{s'}$. The functions $B_\pm(t)$ are correlation functions for the polaron defined in \pref{eq:Bpm_def} and play a role similar to $D^\gtrless(t)$ in the original frame. The structure of $K_{nmkl}^\pm(t)$ is similar to that of the scattering terms in the original frame, but the interpretation is complicated by the fact that we are in the polaron frame.

\subsection{The long-time non-Markovian limit}\label{sec:degree_of_nonmarkov}
The scatterings terms arising from the TCL are time-dependent, giving rise to non-Markovian behavior. In the case of an initial excitation of the system, the duration of the time-dependence is set by the memory depth of the associated reservoir correlation function, $D^\gtrless(t)$ for the original and $B_\pm(t)$ for the polaron frame. This is evident from Eqs. (\ref{eq:Ggl_def}), (\ref{eq:gam12_def}), (\ref{eq:gam13_def}), and (\ref{eq:Knmkl_def_main_paper}) as the time-evolution operator itself for either frame does not decay.

\begin{figure}[ht]
 \centering
 \includegraphics[width=0.45\textwidth]{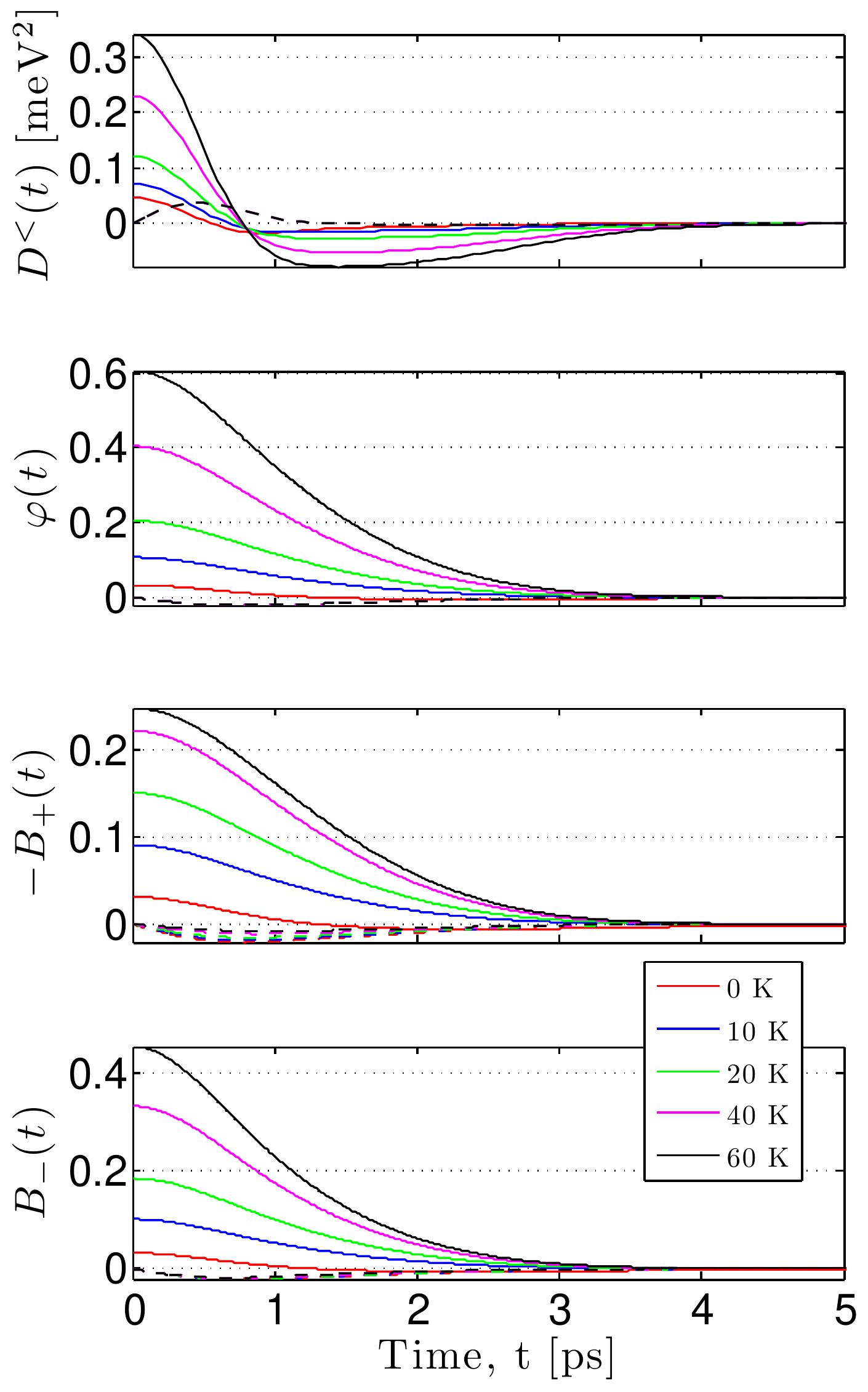}
 \caption{Illustrations of the various correlation functions for the phonon reservoir for fixed material parameters \cite{QD_phonon_parameters} and a range of temperatures. Solid (dashed) lines are for the real (imaginary) part. $D^>(t)$ can be obtained as $D^>(t)=[D^<(t)]^*$.}
\label{fig:phonon_corr_fct_figs}
\end{figure}
In \pref{fig:phonon_corr_fct_figs} we show examples of the various correlation functions for a range of relevant temperatures. The correlation function in the original frame,  $D^\gtrless(t)$, has a temperature independent imaginary part (see \pref{eq:Dgtrless_def}), whereas the real part varies significantly with temperature. The amplitude is smallest and memory depth is largest for low temperatures (the memory depth is extracted from the normalized correlation function, not shown), where an increasing temperature leads to a larger amplitude and smaller memory depth. In the polaron frame the corresponding correlation functions are $B_\pm(t)$, for which both the real and imaginary part are temperature dependent. The amplitude and memory depth behave as in the original frame. For completeness we also show $\varphi(t)$ entering $B_\pm(t)$, see \pref{eq:Bpm_final_result}.

Above we discussed the dependence of the phonon correlation functions on temperature, however other parameters also influence the amplitude and memory depth of the correlation functions. The spatial extent of the QD wavefunction turn out to be important. The phonon coupling matrix element, see \pref{eq:Mk_full_def}, is directly related to the spatial Fourier transform of the absolute square of the wavefunction of the relevant QD state. A small QD will have relatively wide spectrum in $\bs k$-space and thus couple to more phonon modes, causing the corresponding correlation function to decay faster. Conversely, a large QD will have a more narrow spectrum and couple to fewer phonon modes, resulting in a slower decay of the correlation function \cite{Krummheuer2002}. In the following, we keep the size of the QD fixed and will not investigate this further.

From \pref{fig:phonon_corr_fct_figs} we conclude that the time-dependence of the phonon correlation functions and therefore the TCL scattering terms only becomes important within the first few ps of the time evolution. For the time-dependence of the rates to have a significant effect on the dynamics, the RDM has to change significantly within the first few ps after the initial excitation, which is not the case for experimentally relevant parameters. For this reason, we may safely let $t \rightarrow \infty$ in all TCL scattering terms rendering them as constants. While the long-time limit is well justified for studying population decay dynamics, this is not the case for quantities depending sensitively on quantum coherence, e.g., the degree of indistinguishability of single photons \cite{Kaer2012a}.

Taking the $t \rightarrow \infty$ limit in the TCL is sometimes referred to as a Markov approximation \cite{Breuer1999}, whereas the non-Markovian regime is accessed for times smaller than the memory depth of the reservoir. In the case of a memory-less reservoir, the long time limit is exact and does not impose any further approximations. A memory-less reservoir is assumed in the derivation of the famous Lindblad result, see \pref{eq:Lindblad_term}, which is customary referred to as the Markovian limit in the field of cQED.
In our model the reservoir does, however, have memory and we obtain qualitatively different results compared to a Markovian description of the phonon coupling within the Lindblad formalism, even though we take the long time limit in the TCL scattering terms. To distinguish the two qualitatively different descriptions, we will refer to the memory-less (Lindblad) case as the Markovian and the case including memory effects as non-Markovian, even though the $t\rightarrow \infty$ limit has been taken.

\section{Results}\label{sec:results_main}
In this section we present the results obtained from the theory described in the previous sections. In \pref{sec:QD_decay_dyna} we provide a parameter investigation of QD decay dynamics obtained by numerically solving the EOMs in the time-domain and using the polaron frame. We chose the polaron frame in order to obtain the most accurate results. In \pref{sec:anal_approx} we derive analytical expressions for the QD decay rate within both the original and polaron frame. We compare them numerically and discuss the insights that are obtained from their analytical forms.
\subsection{Quantum dot decay dynamics}\label{sec:QD_decay_dyna}
\begin{figure}[ht]
 \centering
 \includegraphics[width=0.45\textwidth]{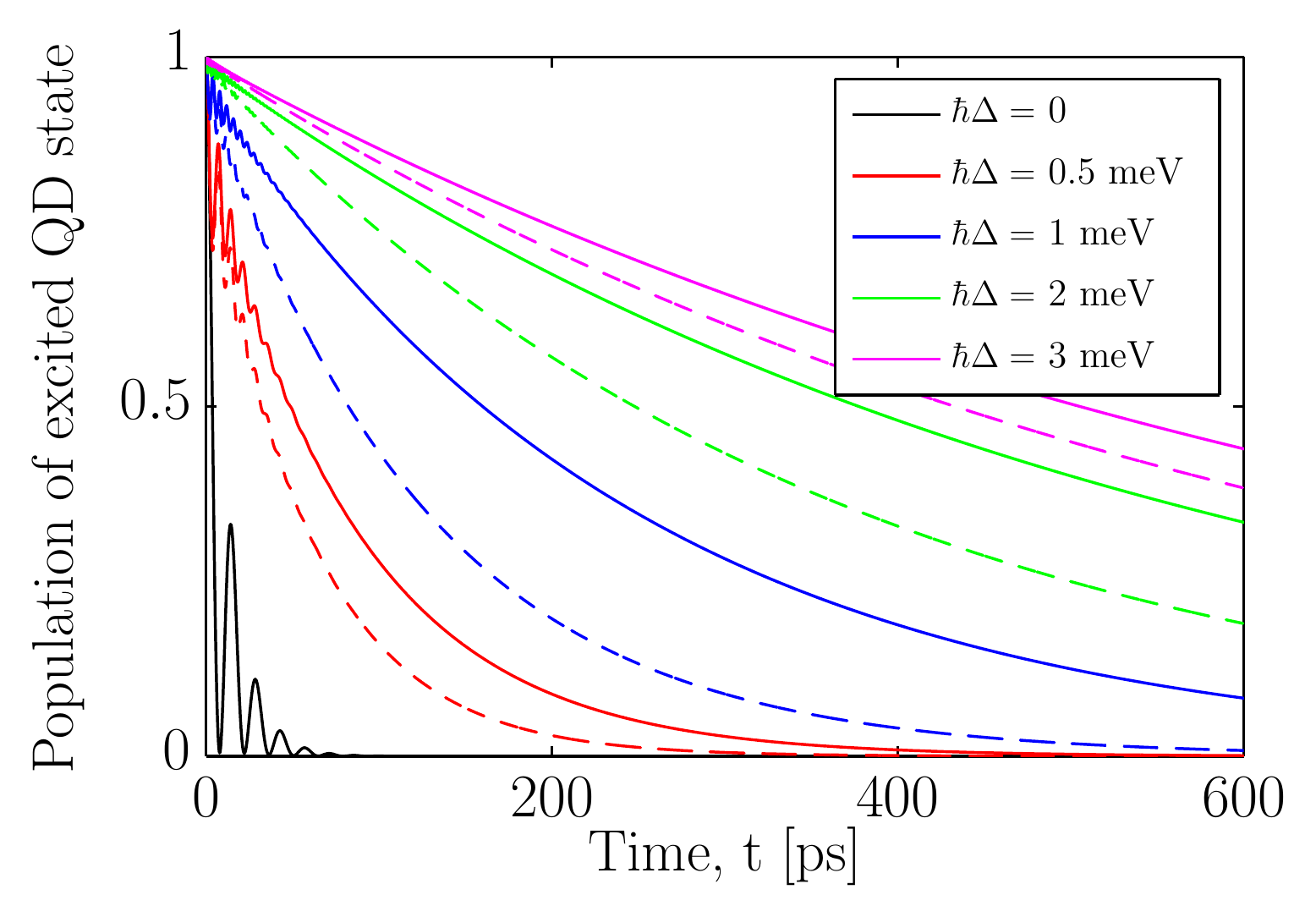}
 \caption{QD decay curves for an initially excited QD, calculated as $\sigg{11}(t)$. The curves are for different signs of the detuning, solid (dashed) is for negative (positive) detuning, defined as $\Delta = \omm{\mrm{eg}}-\omm{\mrm{cav}}$. Parameters: $T=0$ K, $\hbar g = 150~\mu$eV, $\hb \kappa = 100~\mu$eV,  $\Gamma = 1~$ns$^{-1}$, and $\hb \gamma = 0~\mu$eV.}
\label{fig:decaycurves_vs_detuning}
\end{figure}
In \pref{fig:decaycurves_vs_detuning} we show a series of decay curves calculated within the polaron frame for an initially excited QD and compare the results for different signs and values of the detuning \cite{Winger2009a,Hohenester2009c,Kaer2010,Hohenester2010}. The excitation could be due to an optical pulse, resonant with the photon-emitting $\ket g \leftrightarrow \ket e$ transition or higher states of the QD. The chosen parameter values ($g > \kappa,~\Gamma,~\gamma$) places this system well within the so-called strong coupling regime and the temperature has been set to $0$ K to freeze out thermal excitation of phonons.

For the resonant case we observe a very fast decay, and clear Rabi oscillations, indicating the strong coupling regime. For non-zero detuning we observe an asymmetry with respect to the sign of the detuning, which has been predicted theoretically \cite{Kaer2010} and observed experimentally \cite{Winger2009a,Hohenester2009c,Madsen}. The physical origin of the asymmetry is due to spontaneous emission of phonons, while absorption of phonons is unlikely at very low temperatures, which could otherwise restore symmetry. The decay is fastest for positive detuning, as here the initially excited electron may emit a phonon to become resonant with the cavity and decay through it, whereas for negative detuning, the absorption of a phonon is required. It is clearly seen that the asymmetry is strongest for intermediate detuning values, which may be explained by examining the interaction matrix element, see \pref{eq:Mk_full_def}. From the nature of the deformation potential interaction, the matrix element vanishes for small phonon energies becoming proportional to $\sqrt{\om_k}$, while for large energies the form factor imposed by the finite QD wavefunction \cite{Ramsay2010a} causes the matrix element to decay. This gives rise to a maximum in the phonon matrix element, leading to the largest degree of asymmetry.
\begin{figure}[ht]
 \centering
 \includegraphics[width=0.45\textwidth]{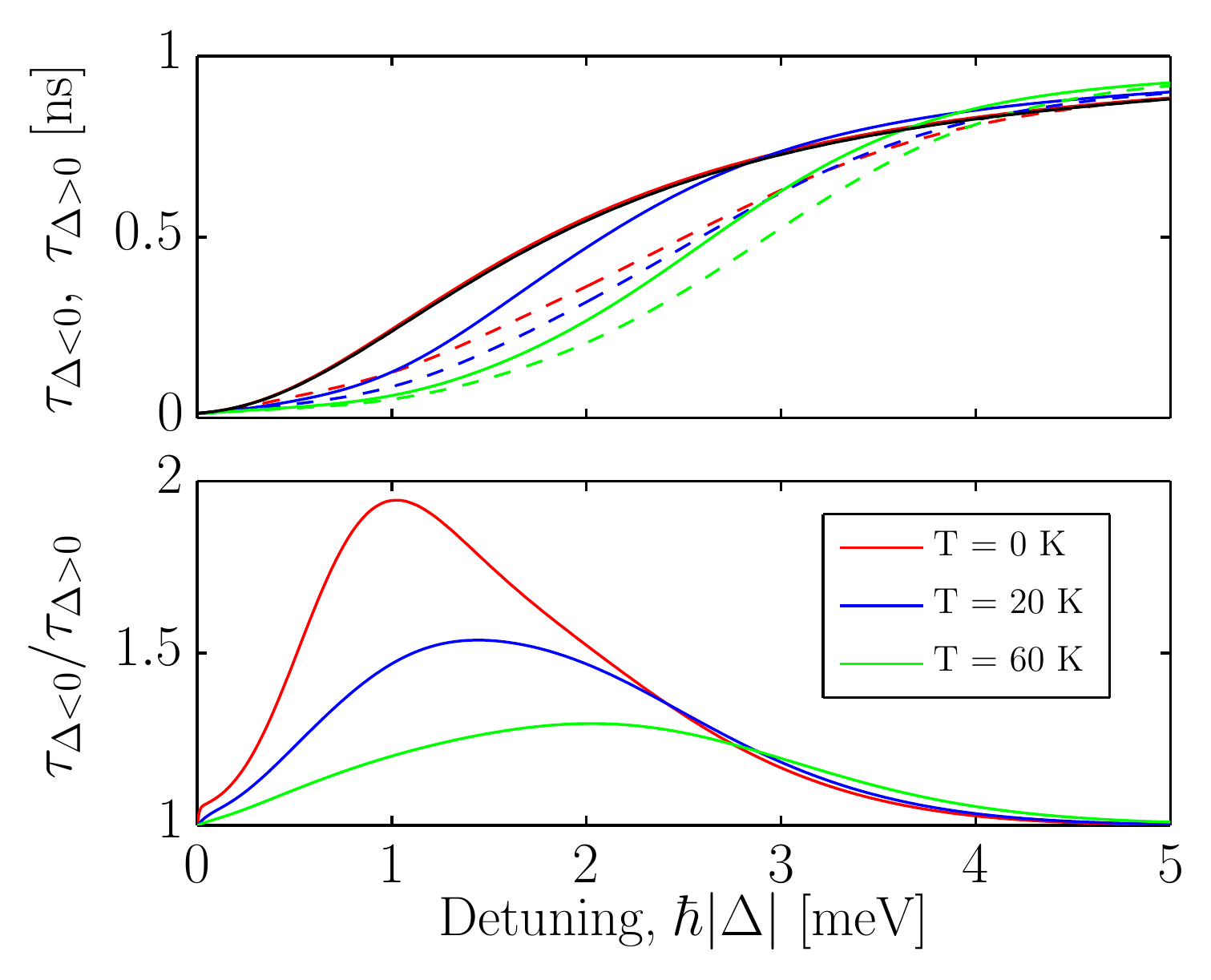}
 \caption{(Top) QD lifetimes for negative, $\tau_{\Delta<0}$, (solid curve) and positive, $\tau_{\Delta>0}$, (dashed curve) detuning at three temperatures for a range detuning values. The black curve is with no phonons in the model. (Bottom) Degree of asymmetry quantified by the ratio between the QD lifetimes for opposite sign of detuning. Parameters are $\hbar g = 150~\mu$eV, $\hb \kappa = 100~\mu$eV, $\Gamma = 1~$ns$^{-1}$, and $\hb \gamma = 0~\mu$eV.}
\label{fig:decayrate_vs_detuning}
\end{figure}

To more systematically quantify the dependence on detuning and the influence of finite temperature on the phonon induced asymmetry, we calculated the degree of asymmetry by taking the ratio between the slow QD lifetime for $\Delta < 0$, $\tau_{\Delta < 0}$, and the faster lifetime obtained for $\Delta > 0$, $\tau_{\Delta > 0}$. The results are presented in \pref{fig:decayrate_vs_detuning} along with the absolute lifetime for both signs of the detuning. The lifetime is obtained by fitting a single exponential to the decay curve obtained from the numerical solution of the model. In the situations where the decay is oscillatory the fitted lifetime thus represents the decaying envelope of the entire curve.
\begin{figure}[ht]
 \centering
 \includegraphics[width=0.45\textwidth]{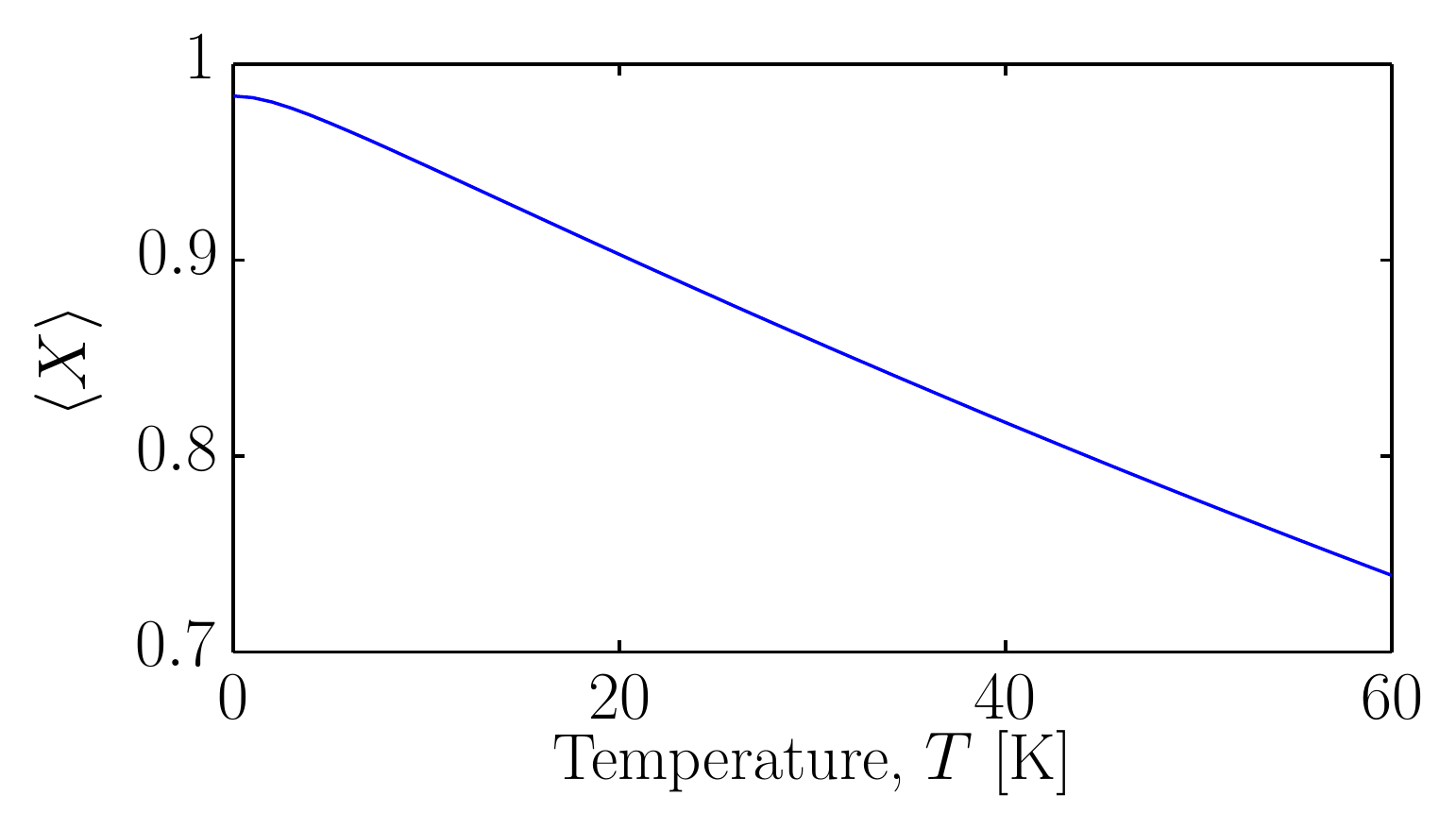}
 \caption{Dependence of $\braket X$, see \pref{eq:X_avg_def}, on temperature.}
\label{fig:renorm_XX}
\end{figure}

For the zero temperature case studied in \pref{fig:decaycurves_vs_detuning}, we observe a degree of asymmetry of almost 2 near a detuning of approximately $1$ meV. The value of the detuning for which the maximum is obtained is determined by the effective size of the QD through the form factor entering the phonon matrix element $M_{\bsk}$ \cite{Nazir2008}. For comparison, we also show the curve with no phonons in the model and which shows that for low temperatures the QD lifetime for $\Delta < 0$ is only very weakly influenced by the phonons. As the temperature is increased the degree of asymmetry decreases. Intriguingly, the QD is seen to decay more slowly at very large detuning as temperature is increased, even though this is basically outside the bandwidth of the phonons. We believe this to be due to the renormalization of $g$ caused by $\braket X$, lowering the effective value of $g$, see \pref{fig:renorm_XX}, where the temperature dependence of $\braket X$ is shown. The smaller asymmetry for higher temperatures is caused by the presence of thermally excited phonons, making it more probable for the electron to absorb a phonon and thereby becoming resonant with the cavity in the case when $\omm{\mrm{cav}} > \omm{\mrm{eg}}$, i.e., $\Delta < 0$.

To illustrate the behavior of the phonons at different temperatures, we calculated the real part of the phonon correlation function \pref{eq:Dgtrless_def} in the frequency domain
\ml{\label{eq:phonon_density}
\mrm{Re}\bS{ D^>(\om) } = \pi \sum_{\bs k} \abs{M^{\bsk}}^2[n_{\bsk}\delta(\om+\omm{\bsk})\\
+\bS{n_{\bsk}+1}\delta(\om-\omm{\bsk})],
}
where the Fourier transform is calculated as $D^>(\omega) = \int_0^\infty dt \e{i(\omega+i0^+) t} D^>(t)$, where $0^+$ is a positive infinitesimal.
The quantity $\mrm{Re}\bS{ D^>(\om) }$ gives information about the phonon modes interacting with the QD for a given temperature and can thus be considered as an effective phonon density. Also, it enters directly into the QD decay rate, as will be demonstrated in \pref{sec:anal_approx}.

\begin{figure}[ht]
 \centering
 \includegraphics[width=0.45\textwidth]{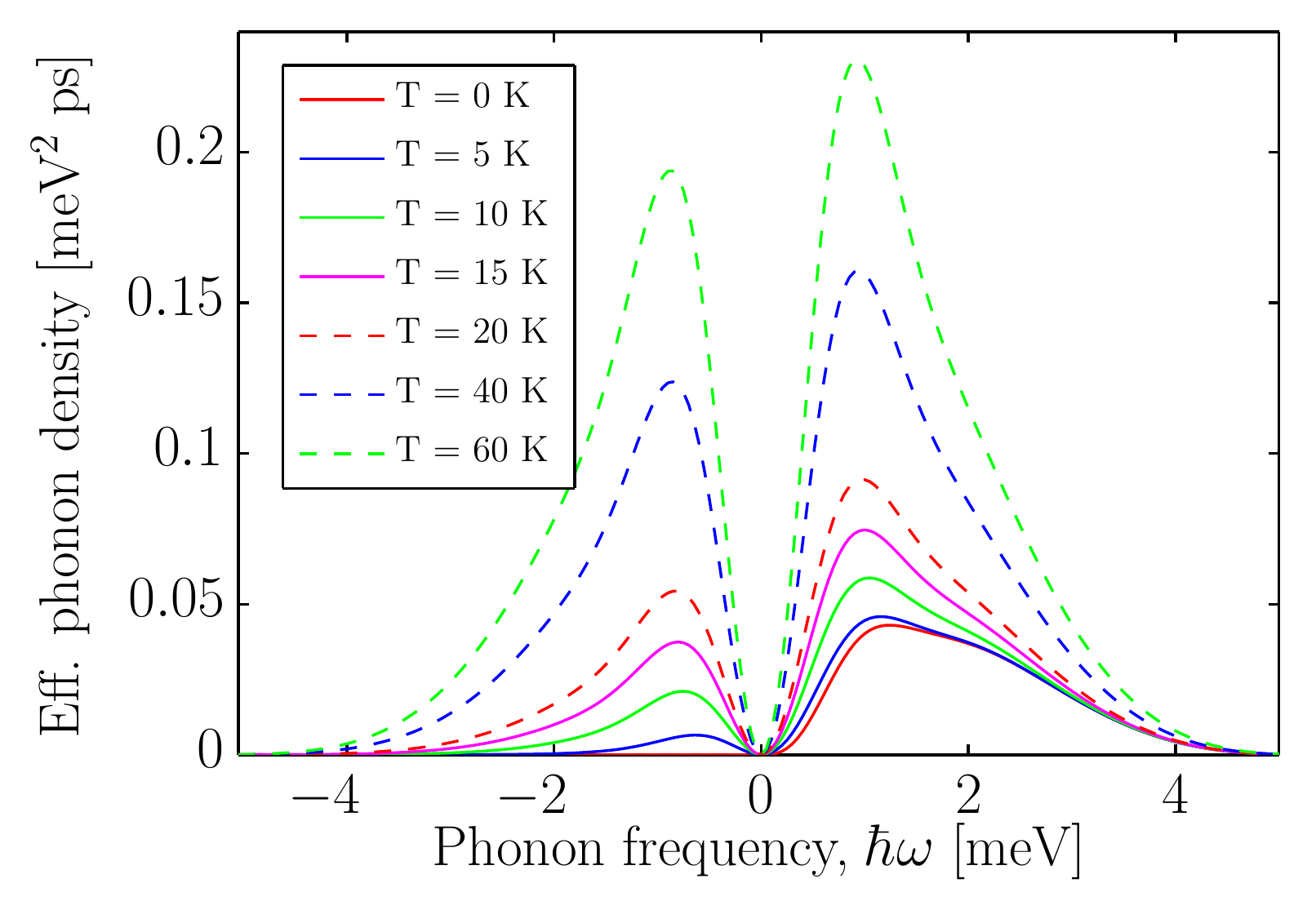}
 \caption{Effective phonon density $\mrm{Re}\bS{ D^>(\om) }$, see \pref{eq:phonon_density}, for a set of temperatures. The weak shoulder visible at low temperatures near $\hb \omega \sim 2~$meV arises due to different localization lengths for the electron in the excited and ground states.}
\label{fig:phonon_density}
\end{figure}
In \pref{fig:phonon_density} we show $\mrm{Re}\bS{ D^>(\om) }$ for a range of temperatures. For zero temperature, no phonons are available for absorption processes, corresponding to negative frequencies in the figure, while the vacuum phonon field reveals its presence through the non-zero density for positive energies. This explains why the asymmetry is largest for zero temperature, as illustrated in \pref{fig:decayrate_vs_detuning}. As the temperature is increased, more and more phonons are being thermally excited and become available for both absorption and stimulated emission processes. The strong asymmetry is no longer present in the effective phonon density, which correlates nicely with the observed behavior of the QD lifetimes.

\begin{figure}[ht]
 \centering
 \includegraphics[width=0.45\textwidth]{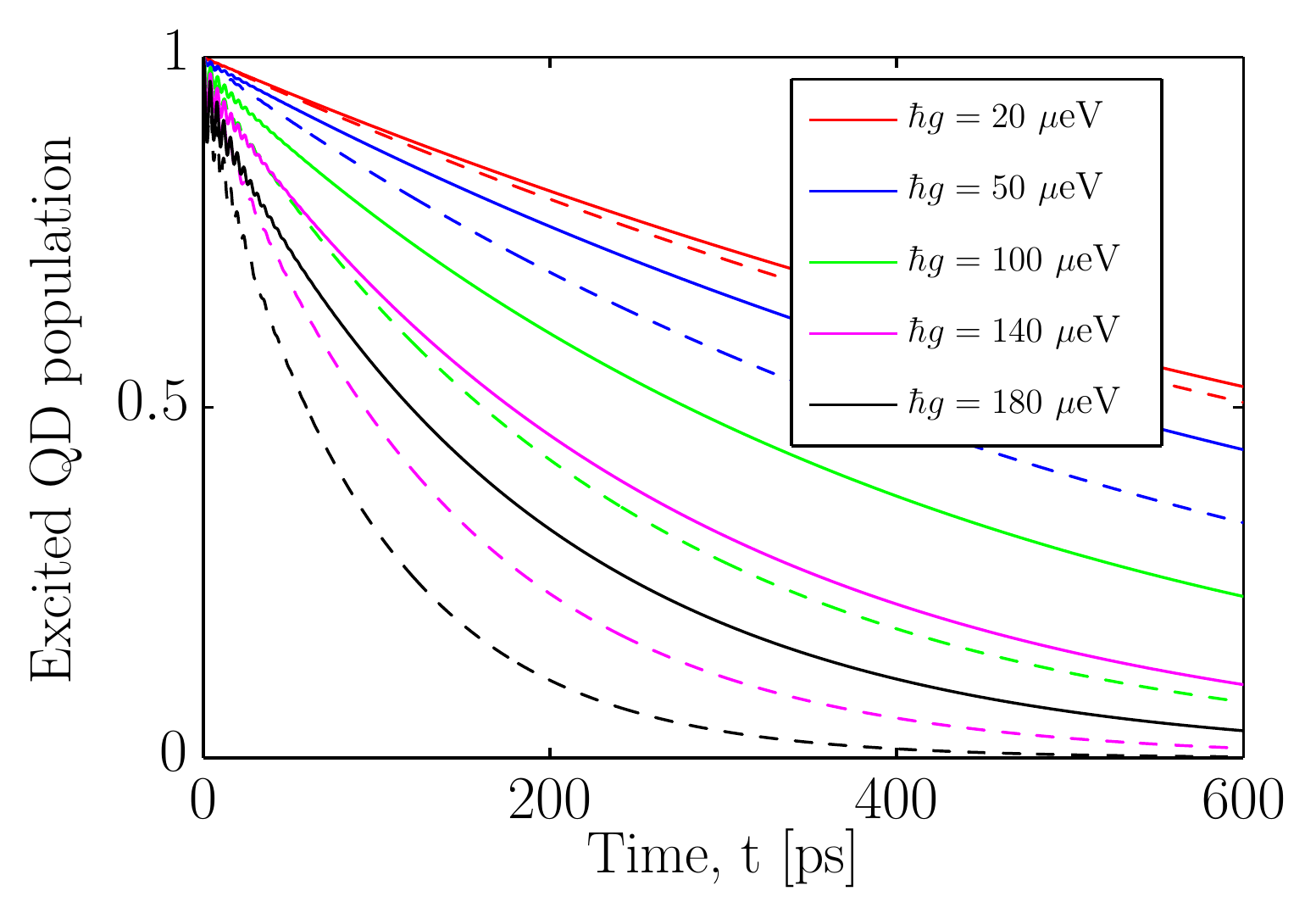}
 \caption{QD decay curves for an initially excited QD. The curves are for different values of the light-matter coupling strength, $g$, with the solid (dashed) lines being for a detuning of -1 (+1) meV. Parameters: $T=0$ K, $\hb \kappa = 100~\mu$eV, $\Gamma = 1~$ns$^{-1}$, and $\hb \gamma = 0~\mu$eV.}
\label{fig:decaycurves_vs_g}
\end{figure}
\begin{figure}[ht]
 \centering
 \includegraphics[width=0.45\textwidth]{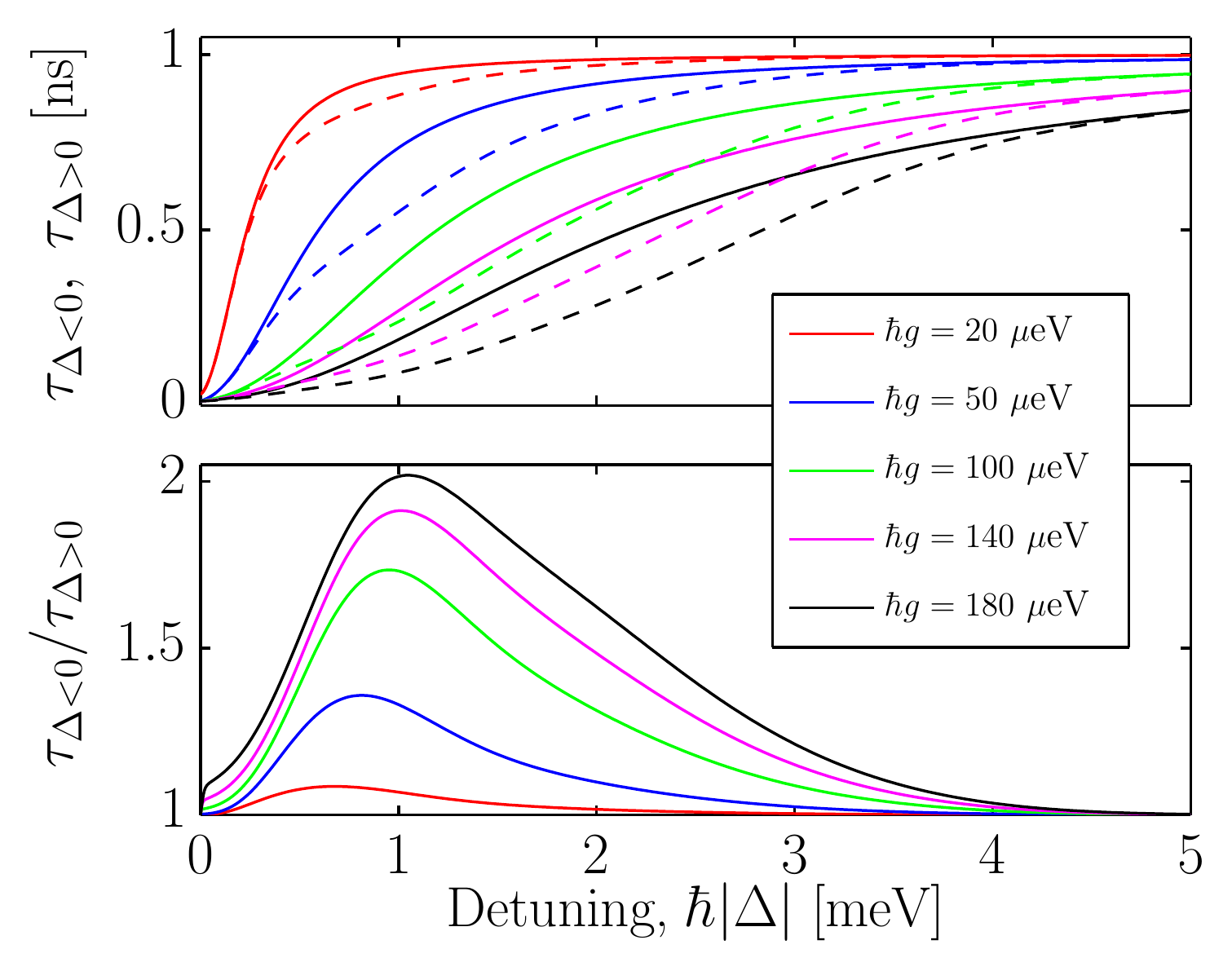}
 \caption{As \pref{fig:decayrate_vs_detuning}, except here the light-matter coupling strength is varied and $T=0~$ K.}
\label{fig:decayrate_vs_g}
\end{figure}
We will now investigate the dependence of the phonon-induced asymmetry on the light-matter coupling strength $g$. In \pref{fig:decaycurves_vs_g} we show decay curves for a QD for both signs of the detuning and vary the light-matter coupling strength from very small values to large values representing current state-of-the-art samples \cite{Winger2009,Reinhard2011}. The temperature is fixed at $0$ K. The first observation is the decrease of lifetime for increasing $g$, consistent with the Purcell effect \cite{Purcell1946}. Furthermore, we also observe an increasing asymmetry between lifetimes for positive and negative detuning values as $g$ is increased.
This trend is seen more clearly in \pref{fig:decayrate_vs_g} where we show the degree of asymmetry as a function of detuning, for varying light-matter coupling strength $g$. It is apparent that one may go from a situation of basically no asymmetry, obtained for a sample in the regime of weak or intermediate coupling strength \cite{Madsen2011}, to more than a factor of 2 in ratio between lifetimes in state-of-the-art samples\cite{Winger2009,Reinhard2011}. This behavior might seem surprising at first, since, as independently of the value of the detuning, the electron has to emit a photon in order to decay to the ground state, regardless of whether a phonon was emitted or absorbed. From this observation one would expect the degree of asymmetry to be independent of $g$, since the Purcell enhancement scales with g, independently of the detuning. The reason for the dependence on $g$ is simple, as will be explained below.

\begin{figure}[ht]
 \centering
 \includegraphics[width=0.45\textwidth]{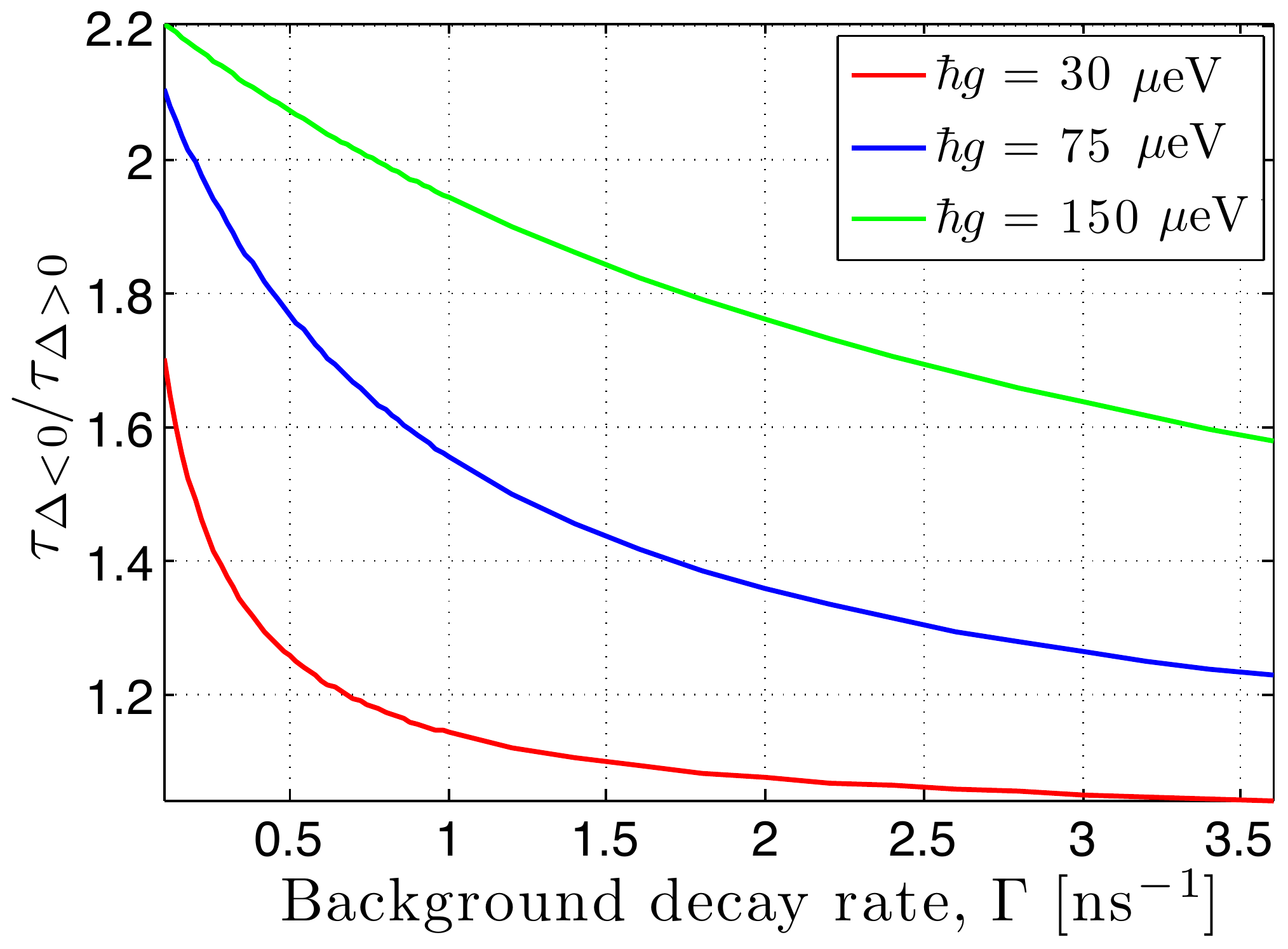}
 \caption{Degree of asymmetry as a function of QD background decay rate. The detuning is fixed at $\hbar\abs{\Delta}=1~$meV, while the light-matter coupling strength is varied. Other parameters are: $T=0$ K, $\hb \kappa = 100~\mu$eV, and $\hb \gamma = 0~\mu$eV.}
\label{fig:vary_background_decay}
\end{figure}
The degree of asymmetry is seen to approach unity in the limit of small light-matter coupling strength, where cavity-mediated effects play a less significant role for the QD decay dynamics. Indeed, in the limit of small $g$ or large $\Delta$, the dominant decay channel for the QD becomes the background decay rate, $\Gamma$, which includes, e.g., decay into radiation modes and non-radiative decay. To illustrate the effect of the background QD decay rate, we show in \pref{fig:vary_background_decay} the degree of asymmetry as a function of $\Gamma$ for a few typical values of the light-matter coupling strength, covering weak, intermediate, and strong coupling. For a typical weak coupling sample, $\hbar g=30~\mu$eV, a noticeable asymmetry is only visible for very small $\Gamma$, corresponding to cavities where radiation modes are strongly suppressed such as photonic crystal cavities. The asymmetry disappears as the phonon contributions become dominated by the background decay rate. On the other hand, for a sample well within the strong coupling regime, $\hbar g=150~\mu$eV, a significant asymmetry should be observable for basically all values of the background decay rate.

\subsection{Approximate analytical expressions}\label{sec:anal_approx}
While the results from the previous section are numerically exact solutions for the dynamics, more physical insight can be gained through approximate analytical expressions for the QD decay rates. In the limit of large detuning, $\Delta \gg g$, such expressions can be obtained in both the original and polaron frame. This is possible as we can adiabatically eliminate the involved polarizations, and the time evolution operator, $U(t)$, may be expanded to a low order in the quantity $g/\Delta$, see \pref{app:anal_rates} for details.

In the original frame we obtain the following expression for the total QD decay rate
\ml{\label{eq:purcell_orig}
\Gamma_\mrm{tot} = \Gamma \\+ 2g^2\frac{\gamma_\mrm{tot}}{\gamma^2_\mrm{tot}+\Delta^2}\bT{ 1 + \frac{1}{\hb^2\gamma_\mrm{tot}}\mrm{Re}\bS{D^>(\om = \Delta)} },
}
and for the polaron frame we obtain
\ml{\label{eq:purcell_pol}
\Gamma'_\mrm{tot} = \Gamma \\+ 2[g\braket{X}]^2\frac{\gamma_\mrm{tot}}{\gamma^2_\mrm{tot}+\Delta^2}+2g^2\mrm{Re}[B_-(\om=\Delta)],
}
where the total dephasing rate is defined as
\al{
\gamma_\mrm{tot} = \frac 12(\kappa+\Gamma)+\gamma.
}
In Eqs. (\ref{eq:purcell_orig}) and (\ref{eq:purcell_pol}) the Fourier transform is calculated as $f(\omega) = \int_0^\infty dt \e{i(\omega+i0^+) t} f(t)$, where $0^+$ is a positive infinitesimal.

\begin{figure}[ht]
 \centering
 \includegraphics[width=0.45\textwidth]{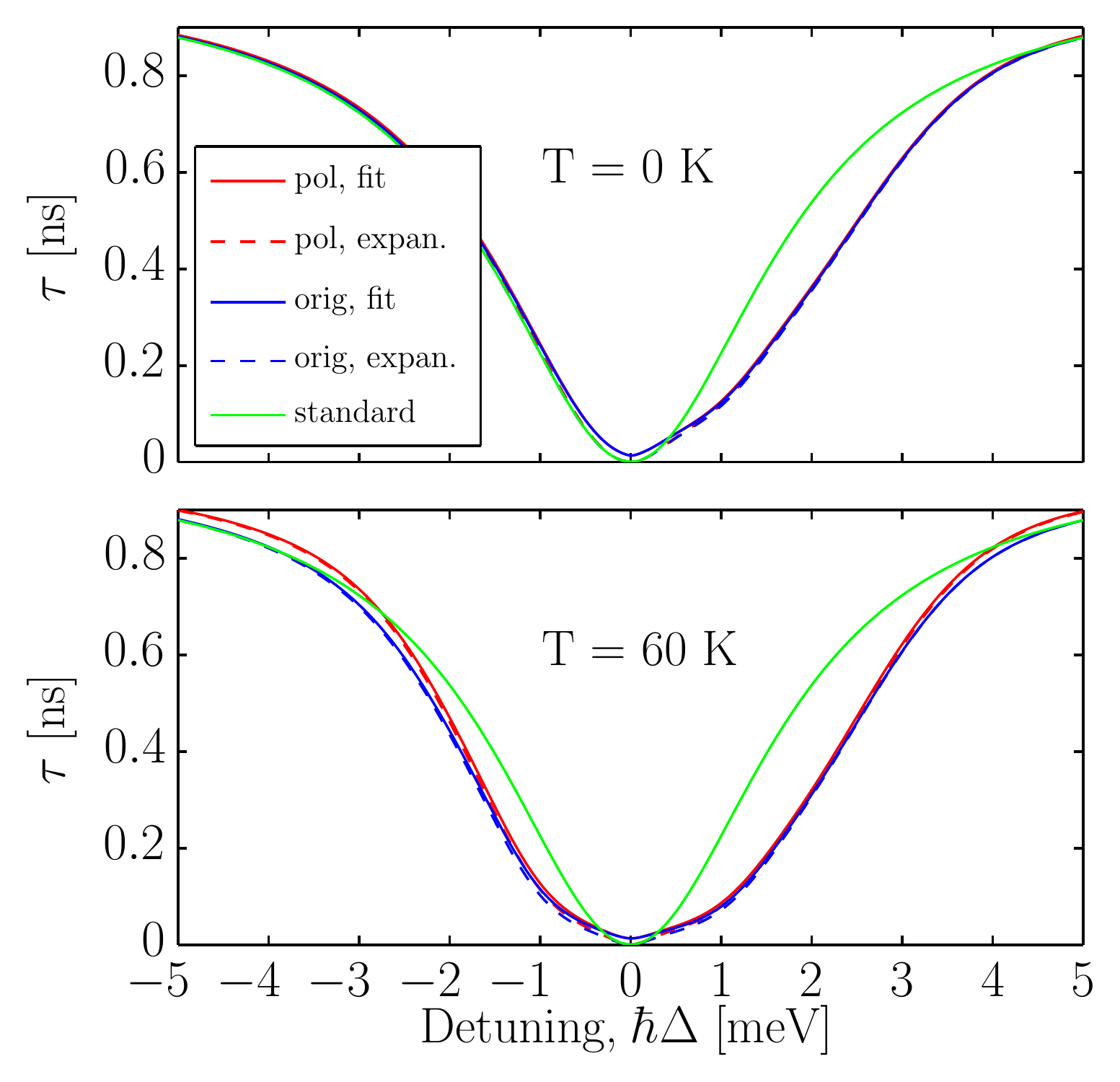}
 \caption{Comparison of QD lifetimes obtained through the approximate rates (dashed curves) in the original (blue), \pref{eq:purcell_orig}, and polaron frame (red), \pref{eq:purcell_pol}, and a single exponential fit (solid curves) to the numerically exact solution. We also show the result when phonons are not included in the model (green). Parameters are: $\hb \kappa = 100~\mu$eV, $\hb g = 150~\mu$eV, $\Gamma = 1~$ns$^{-1}$, and $\hb \gamma = 0~\mu$eV.}
\label{fig:exact_vs_approx_kap_100}
\end{figure}
\begin{figure}[ht]
 \centering
 \includegraphics[width=0.45\textwidth]{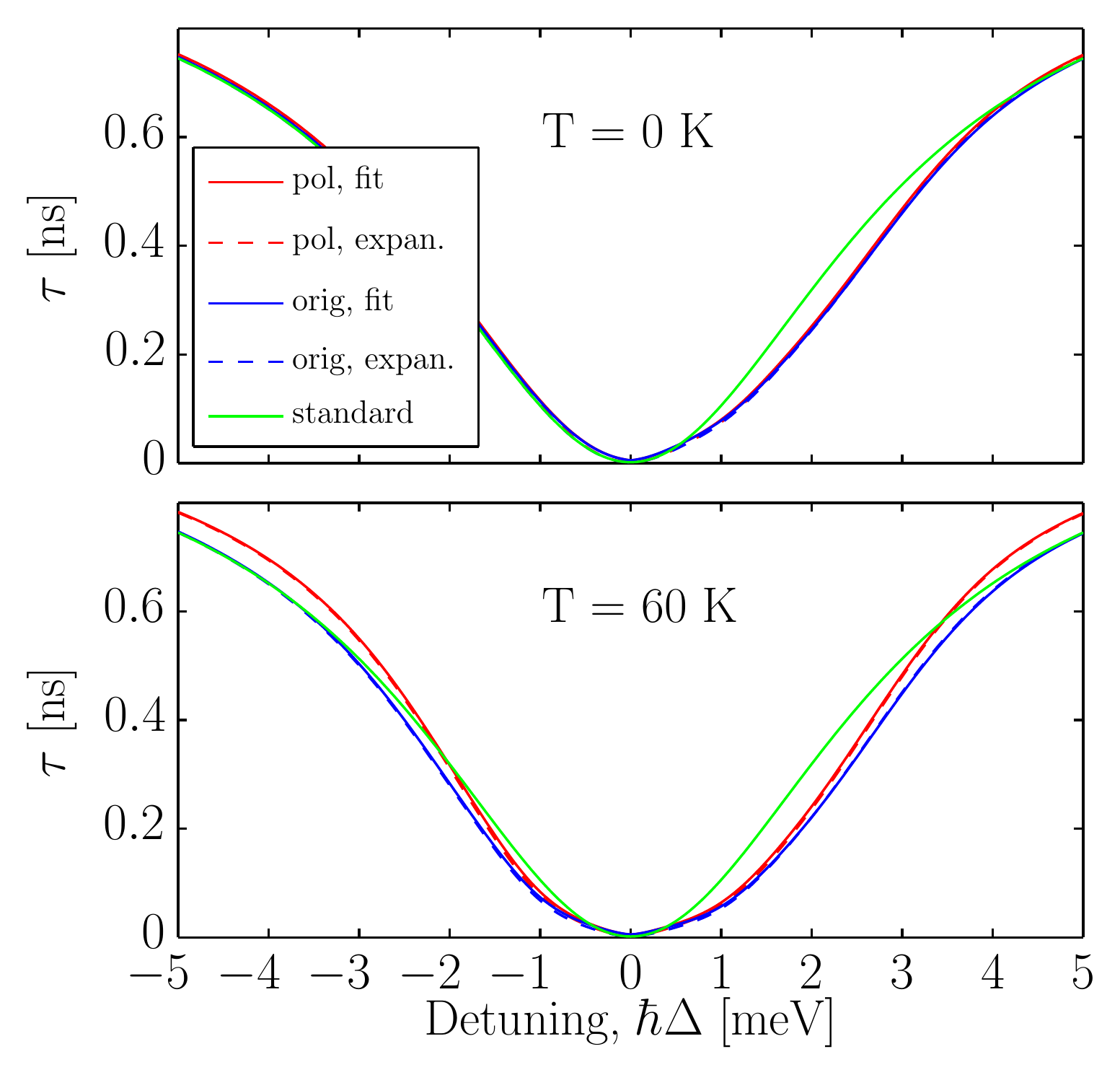}
 \caption{As in \pref{fig:exact_vs_approx_kap_100}, except that $\hb \kappa = 250~\mu$eV.}
\label{fig:exact_vs_approx_kap_250}
\end{figure}
In Figs. \ref{fig:exact_vs_approx_kap_100} and \ref{fig:exact_vs_approx_kap_250} we compare the QD lifetime ($\tau=1/\Gamma_\mrm{tot}$) calculated from the approximate expressions with single exponential fits to the numerically exact solutions, for two typical sets of parameters. For all but very small detuning values, the approximate expressions compare very well to the corresponding numerical fits. The strong asymmetry at low temperatures, as well as the more symmetric decay rates at elevated temperatures, are well captured by the approximate expressions. At high temperatures, we observe significant deviation between the results in the original and the polaron frame. This is expected as only the polaron frame takes into account multi-phonon effects that become increasingly important at elevated temperatures \cite{Hohenester2010,McCutcheon2010}.

The expression for the decay rate in the original frame, \pref{eq:purcell_orig}, has a form very suitable for interpretation. In addition to the background QD decay rate $\Gamma$, there are two contributions. The first contribution accounts for the direct decay of the QD through the cavity by emission of a photon, with the total dephasing rate $\gamma_\mrm{tot}$ including a Lindblad pure dephasing rate $\gamma$ \cite{Auffeves2010}. This gives rise to the familiar symmetric dependence on the detuning, see the green curve in Figs. \ref{fig:exact_vs_approx_kap_100} and \ref{fig:exact_vs_approx_kap_250}. However, the second contribution goes beyond the standard models of cQED by depending on the effective phonon density $\mrm{Re}[D^>(\om = \Delta)]$ evaluated at the QD-cavity detuning, see \pref{eq:phonon_density} and \pref{fig:phonon_density}. Thus, the phonon-assisted QD decay simultaneously depends on the cavity, through the Purcell rate prefactor, and on the availability of phonons that couple to the QD at the given QD-cavity detuning. Loosely, one can think of the second contribution as a product between the effective photon and phonon densities available for both spontaneous and stimulated processes.

\begin{figure}[ht]
 \centering
 \includegraphics[width=0.5\textwidth]{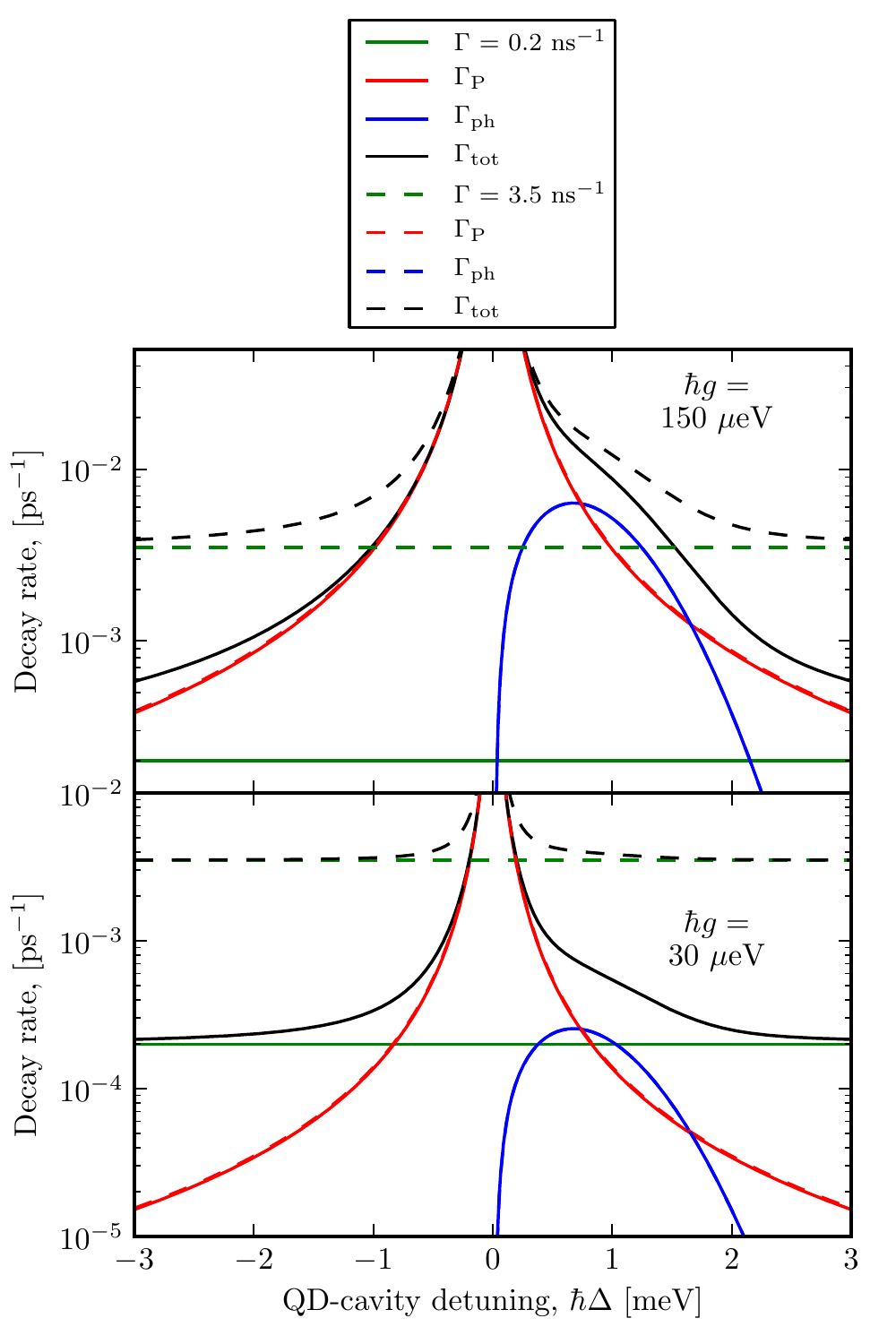}
 \caption{Contributions to the total QD decay rates given in \pref{eq:orig_anal_rate_split}. Note that for these figures a spherical QD model has been employed, using a harmonic confinement length of $5~$nm. Parameters: $\hbar\kappa=100~\mu$eV, $\hbar\gamma=0~\mu$eV, and $T=0~$K.}
\label{fig:anal_QD_rate}
\end{figure}
Based on the analytical expression for the QD decay rate in the original frame, \pref{eq:purcell_orig}, we can provide a more physically transparent discussion of the dependence on $g$ and $\Gamma$ of the degree of asymmetry discussed in Figs. \ref{fig:decaycurves_vs_g}, \ref{fig:decayrate_vs_g}, and \ref{fig:vary_background_decay}.
We begin by formally dividing the total QD decay rate into the three contributions discussed above
\al{\label{eq:orig_anal_rate_split}
\Gamma_\mrm{tot} &= \Gamma + \Gamma_\mrm{P}+\Gamma_\mrm{ph},\\
\Gamma_\mrm{P}&=2g^2\frac{\gamma_\mrm{tot}}{\gamma^2_\mrm{tot}+\Delta^2},\\\label{eq:orig_phon_ass}
\Gamma_\mrm{ph}&=2g^2\frac{\hb^{-2}}{\gamma^2_\mrm{tot}+\Delta^2}\ree\bS{D^>(\om = \Delta)},
}
where $\Gamma$ is the background decay rate, $\Gamma_\mrm{P}$ is the usual Purcell enhanced rate, and $\Gamma_\mrm{ph}$ is the rate containing the phonon contribution and can be thought of as a phonon-assisted Purcell enhanced rate.
With reference to Fig. \ref{fig:vary_background_decay} we show in Fig. \ref{fig:anal_QD_rate} the three contributions to $\Gamma_\mrm{tot}$ for two values, one small and one large, of $\Gamma$ and $g$, as a function of detuning.
For both values of the QD-cavity coupling, we observe that neither the bare Purcell rate nor the phonon-assisted rate are affected much by going from the small background decay rate, $\Gamma=0.2~$ns$^{-1}$, to the larger background rate, $\Gamma=3.5~$ns$^{-1}$.
Close to resonance, also the total decay rate appears rather independent of the magnitude of the background as it is completely dominated by the bare Purcell enhanced rate.
However, this picture changes dramatically once we increase the detuning and the contribution from the bare Purcell rate becomes comparable to the two other contributions.
In the case of the large background rate and small QD-cavity coupling, $\hbar g=30~\mu$eV, the constant background dominates over the phonon-assisted rate, $\Gamma_\mrm{ph}$, and hardly any phonon-induced asymmetry is observed.
Referring to Fig. \ref{fig:vary_background_decay} this situation corresponds to a typical micropillar cavity in the weak coupling regime.
If we now decrease the background rate to a lower value, corresponding to a typical photonic crystal cavity in the weak coupling regime [Fig. \ref{fig:vary_background_decay}], the background and the phonon-assisted contributions become comparable and the degree of asymmetry consequently rises.
This illustrates that one may enter a regime, where phonon-induced spectral asymmetries become significant, by changing the background decay, a parameter which is often thought of as being of minor importance and with trivial physical implications.
Increasing the QD-cavity coupling to values typically found in the strong coupling regime, $\hbar g=150~\mu$eV, we significantly increase both the bare and the phonon-assisted Purcell enhanced rates.
For both values of the background rate, a clear asymmetry in the total QD decay rate is now observed, owing to the fact that the constant and symmetric background rate no longer masks the phonon-assisted decay rates.

The approximate expression in the polaron frame, see \pref{eq:purcell_pol}, is not as straightforward to interpret as the expression in the original frame. The background decay $\Gamma$ enters in the same fashion and we also observe a term similar to the one representing decay directly through the cavity in the original frame. However, in contrast, the quantity $\braket X$ only enters the polaron frame, where it plays the role of renormalizing the light-matter coupling strength to a smaller value. The dependence of $\braket X$ on temperature is shown in \pref{fig:renorm_XX}, where it is seen that the renormalization can be quite significant. The last term involves the spectral properties of the phonons, through the Fourier transform of the correlation function $B_-(t)$
\ml{
2g^2\mrm{Re}[B_-(\om=\Delta)]=\\2g^2\braket{X}^2\mrm{Re}\bS{ \int_0^\infty dt \e{i\Delta t} \bT{\e{\varphi(t)}-1}},
}
where $\varphi(t)$ is defined in \pref{eq:phi_def} and plays the role of a phonon-assisted QD decay rate analogous to \pref{eq:orig_phon_ass} in the original frame. As $B_-(t)$ contains $\braket{X}^2$ as a factor, $g$ is renormalized by $\braket X$ everywhere it appears. This is not the case for other cQED models also employing the polaron transformation \cite{Hohenester2010}. The same formula has recently been independently derived and discussed by Roy and Hughes in Ref. \onlinecite{Roy2011b}.

The remaining part involving the Fourier integral over $\mrm{exp} [ \varphi(t) ]-1$ is harder to interpret than the corresponding expression for $D^>(\omega)$ in the original frame. Even though $\varphi(t)$ and $D^>(t)$ appear rather similar, compare \pref{eq:Dgtrless_def} and \pref{eq:phi_def}, Re$[D^>(\omega)]$ directly reflects the effective spectral features of the phonon reservoir. Also, in the original frame, $D^>(\omega)$ carries the familiar Lorentzian-style denominator of the cavity lineshape, which is missing in the polaron frame. Mathematically, the Lorentzian denominator appears in the expression since the phonon induced term enters via a polarization, whereas in the polaron frame, it enters directly as a lifetime. Despite the fact that they superficially look rather different, their numerical values compare very well, especially for low temperatures, as evidenced in Figs. \ref{fig:exact_vs_approx_kap_100} and \ref{fig:exact_vs_approx_kap_250}.

%




\section{Summary and conclusion}\label{sec:summary_conclusion}
In summary, we have presented a theory for coupled QD-cavity systems including the interaction with phonons and illustrated the importance of the phonon interaction for the QD decay dynamics.

Furthermore, we have provided a detailed account of the theory used in recent studies \cite{Kaer2010,Madsen,Nysteen}, which is based on a second order expansion in the phonon coupling, while accounting for the polaritonic nature of the QD-cavity to all orders. It was shown that it is essential to include the polaritonic nature in the interaction, when describing non-Markovian phonon reservoirs.

For elevated temperatures, multi-phonon effects are expected to play an important role. To study the influence of phonons in this regime, we included a theory based on the so-called polaron transformation, which takes certain phonon processes into account to infinite order, while still maintaining important polaritonic aspects of the QD-cavity system.

Using the polaron theory, an extensive investigation of the parameter dependence of the QD decay dynamics was carried out for experimentally relevant regimes. An asymmetric detuning-dependence of the QD lifetime was observed, where a positive detuning, $\om_\mrm{eg}>\om_\mrm{cav}$, yielded a significantly faster decay compared to negative detuning, $\om_\mrm{eg}<\om_\mrm{cav}$. The faster decay observed for positive detuning reflects that the QD may emit a photon by the simultaneous emission of a phonon, thereby overcoming the energy mismatch. Conversely, for negative detuning, absorption of a phonon is required to bridge the gap in energy, but at low temperatures phonon absorption is very unlikely. As the temperature is increased, the asymmetry gradually disappears, due to the availability of phonon absorption processes. Apart from inducing spectral asymmetries, the interaction with phonons also gives rise to a significantly increased bandwidth of the QD-cavity interaction. It greatly extents the bandwidth beyond that imposed by the cavity linewidth normally thought to be the limiting factor, relaxing the resonant nature of many cQED phenomena.

We also provide a simple explanation for the lack of experimental observations of phonon-induced asymmetries in QD decay curves until recently \cite{Hohenester2009c,Madsen,Winger2009a}. We showed how the background decay rate of the QD, often considered insignificant compared to other loss channels, plays a surprisingly important role in observing phonon effects for non-zero detuning. Phonon effects are strongest at relatively large detunings, $1-2$ meV in our case, which typically spans many cavity linewidths of $0.05-0.3$ meV, and thus the effect of the cavity is usually small at these detunings. In order for cavity-mediated effects, such as the phonon asymmetry, to remain significant either a small background decay or a large light-matter coupling strength is needed. Both of these requirements demand high quality samples, which have only become available recently.

To provide further insight into the physics, we derived approximate analytical expressions for the total QD decay rate, which distills the essential ingredients added by the phonon interaction to well-known results from cQED. The power and accuracy of these expressions has recently been demonstrated experimentally and the effective phonon density has been experimentally extracted \cite{Madsen}.

\begin{acknowledgments}
The authors would like to thank A. Grodecka-Grad, C. Roy, and A. Nysteen for helpful discussions. The Center for Nanostructured Graphene is sponsored by the Danish National Research Foundation.
\end{acknowledgments}

\appendix

\section{Equation of motion for reduced density matrix}\label{app:EOM_RDM}
In this appendix we derive the equation of motion for the reduced density matrix of the QD-cavity system, which interacts with a large bosonic reservoir \cite{Breuer2002,Carmichael1999}.

We start by defining the total Hamiltonian
\al{\label{eq:HSR_uber}
H(t) = H_\mrm S (t) + H_\mrm R + H_\mrm{SR} = H_0(t)+H_\mrm{SR},
}
where $H_\mrm S (t)$ is the, possibly time-dependent, Hamiltonian for the system of interest, $H_\mrm R$ is the Hamiltonian for the reservoir, and $H_\mrm{SR}$ is the interaction between the two subsystems. For notational simplicity, we have introduced $H_0(t)$ as the sum of the free contributions.

The time evolution of the total density matrix, $\chi(t)$, is governed by the following equation in the Schr\"{o}dinger picture
\al{\label{eq:totalDM_EOM}
i\hb\pd t \chi(t) =  \bS{H(t),\chi(t)},
}
where $H(t)$ is the Hamiltonian defined in \pref{eq:HSR_uber}. We transform into the interaction picture with respect to $H_\mrm S (t) + H_\mrm R$, to facilitate a perturbation expansion in orders of the interaction $H_\mrm{SR}$. The transformation operator $U_{H_0(t)}(t,t_0)$ satisfies the Schr\"{o}dinger equation
\al{\notag
i\hb \pd t U_{H_0(t)}(t,t_0) &= \bT{H_\mrm S (t) + H_\mrm R} U_{H_0(t)}(t,t_0)\\
 &= H_0(t) U_{H_0(t)}(t,t_0),
}
where $t_0$ is the initial time, and $U_{H_0(t)}(t_0,t_0)=I$, with $I$ being the identity operator. $U_{H_0(t)}(t,t_0)$ may be formally integrated, and due to the allowed time-dependence of the system Hamiltonian, we end up with the time-ordered expression
\al{
U_{H_0(t)}(t,t_0) = T\bT{ \exp \bP{ -i\hbm \int^t_{t_0} dt' H_0 (t') } },
}
with $T$ being the time-ordering operator. The interaction picture representation of the total density matrix is defined as
\al{
\ti \chi(t) = \hc U_{H_0(t)}(t,t_0) \chi (t) U_{H_0(t)}(t,t_0),
}
which leads to the following equation of motion for $\ti \chi(t)$
\al{\label{eq:totalDM_EOM_IP}
i\hb \pd t \ti\chi (t) = \bS{\ti H_\mrm{SR}(t),\ti \chi (t)}.
}
This equation can be formally integrated
\al{
\ti\chi (t) = \ti\chi (t_0)-i\hbm \int_{t_0}^t dt' \bS{\ti H_\mrm{SR}(t'),\ti \chi (t')}.
}
By inserting this expression into the right hand side of \pref{eq:totalDM_EOM_IP} and tracing over the reservoir degrees of freedom, we obtain a formally exact equation for the reduced density matrix of the system
\ml{\label{eq:mother_exact_DM_eom}
i\hb \pd t \ti\rho (t) = \Tr_\mrm R\bT{\bS{\ti H_\mrm{SR}(t),\ti \chi (t_0)}} \\
-i\hbm \int_{t_0}^t dt'\Tr_\mrm R\bT{\bS{\ti H_\mrm{SR}(t),\bS{\ti H_\mrm{SR}(t'),\ti \chi (t')}}},
}
where $\rho(t) = \Tr_\mrm R\bT{\chi (t)}$ and $\Tr_\mrm R \bT{ \cdots }$ denotes the trace operation with respect to the reservoir degrees of freedom. To proceed further, we need to start invoking approximations. The first approximation, known as the Born approximation, assumes that the total density matrix on the right hand side of \pref{eq:mother_exact_DM_eom} factorizes at all times, and especially at the initial time, hence
\al{\label{eq:Born_approx}
\chi(t) = \rho(t)R_0 \Leftrightarrow \ti\chi(t) = \ti\rho(t)\ti R_0 = \ti\rho(t) R_0,
}
where $R_0=\ti R_0$ is the density matrix for the reservoir, assumed to remain in a thermal state at all times and hence being time-independent. This approximation is expected to hold for weak interaction between the system and reservoir. Motivated by the specific physical situation considered, we shall assume that $H_\mrm{SR}$ is written on the following form
\al{\label{eq:HSR_specific_form}
H_\mrm{SR} = \sum_{\nu\nu'}P_{\nu\nu'}B_{\nu\nu'},
}
where $P_{\nu\nu'}$ is a pure system operator and $B_{\nu\nu'}$ is a pure reservoir operator. We assume that $B_{\nu\nu'}$ has the following property
\al{\label{eq:Bres_condition}
\Tr_\mrm R \bT{R_0 B_{\nu\nu'} }= \braket{B_{\nu\nu'}}_{0} = 0.
}
If we now use eqs. (\ref{eq:Born_approx}) and (\ref{eq:Bres_condition}) in \pref{eq:mother_exact_DM_eom} we arrive at
\al{\label{eq:mother_DM_eom_1}
\pd t \ti\rho (t) = -\hb^{-2} \int_{t_0}^t dt'\Tr_\mrm R\bT{\bS{\ti H_\mrm{SR}(t),\bS{\ti H_\mrm{SR}(t'),\ti\rho(t') R_0}}},
}
which completes the formal derivation of the equation of motion for the reduced density matrix.

To use the specific form of the interaction Hamiltonian, \pref{eq:HSR_specific_form}, we insert this into the above, expand the commutators, and rearrange the position of the $B_{\nu\nu'}$'s with respect $R_0$ to obtain well-defined expectation values over the reservoir operators. Performing these steps yields
\begin{widetext}
\ml{\label{eq:mother_DM_eom_2}
\pd t \ti\rho (t) =
-\hb^{-2} \int_{t_0}^t dt' \sum_{\nu_1\nu_2\nu'_1\nu'_2} \Big\{ \\
\bT{ \ti P_{\nu_1\nu_2}(t) \ti P_{\nu'_1\nu'_2}(t') \ti \rho(t') - \ti P_{\nu'_1\nu'_2}(t') \ti \rho(t') \ti P_{\nu_1\nu_2}(t)  }\braket{\ti B_{\nu_1\nu_2}(t) \ti B_{\nu'_1\nu'_2}(t')}_0 \\
+\bT{ \ti \rho(t') \ti P_{\nu'_1\nu'_2}(t') \ti P_{\nu_1\nu_2}(t) - \ti P_{\nu_1\nu_2}(t)\ti \rho(t') \ti P_{\nu'_1\nu'_2}(t')  } \braket{\ti B_{\nu'_1\nu'_2}(t') \ti B_{\nu_1\nu_2}(t)}_0  \Big\}.
}
\end{widetext}
In its present form \pref{eq:mother_DM_eom_2} contains a memory integral with $\ti\rho(t')$ as an integrand, therefore the time evolution depends on the past state of the system and therefore non-Markovian. However, it is well-known that a non-Markovian description may also be obtained in a fully time local theory, where the time evolution only depends on the present state of the system, but with time-dependent coefficients arising from the reservoir interaction. One example of such a theory is the timeconvolution-less approach (TCL) \cite{Breuer2002,Breuer1999,Kaer2010,Nazir2008,Rozbicki2008}. In fact, if one makes the replacement $\ti \rho(t') \rightarrow \ti \rho(t)$ in \pref{eq:mother_DM_eom_2} the formal second order result in the TCL is obtained, which still describes a non-Markovian time-evolution. However, it is essential that this replacement is made within the interaction picture, where the only relevant time scale is the assumed slow time scale induced by the interaction with the reservoir \cite{Brandes2004}. We will present the result for both the time-local and memory theory below.

In \pref{eq:mother_DM_eom_2} the time-evolution of the operators is only governed by the free Hamiltonian of the respective subsystem. Thus only the time-evolution operator for the system
\al{
U(t,t_0) = T\bT{ \exp \bP{ -i\hbm \int^t_{t_0} dt' H_\mrm S (t') } },
}
should be used when transforming \pref{eq:mother_DM_eom_2} back to the Schr\"{o}dinger picture. We obtain
\al{\label{eq:reduced_DM_eom}
\pd t \rho(t) =-i\hbm\bS{H_\mrm S(t),\rho(t)}+S(t),
}
where we introduced the reservoir induced scattering term defined as
\al{\label{eq:scatt_term_def}
S(t)=U(t,t_0)[\pd t \ti\rho (t)]\hc U(t,t_0).
}
By employing relations such as
\al{
U(t,t_0) \ti\rho(t') \hc U(t,t_0) = U(t,t') \rho(t')\hc U(t,t'),
}
and
\al{
 U(t,t_0) \ti P_{\nu_1\nu_2}(t') \hc U(t,t_0) = U(t,t') P_{\nu_1\nu_2}\hc U(t,t'),
}
we may derive the final form of the reservoir induced scattering $S(t)$ term, defined in \pref{eq:scatt_term_def}, for both the time-local and memory version described above. The scattering term with memory becomes

\begin{widetext}
\ml{\label{eq:scatt_memory}
S_\mrm{MEM} (t) =
-\hb^{-2} \int_{t_0}^t dt' \sum_{\nu_1\nu_2\nu'_1\nu'_2} \Big\{ \\
\bT{ P_{\nu_1\nu_2} U(t,t') P_{\nu'_1\nu'_2} \rho(t') \hc U(t,t') - U(t,t') P_{\nu'_1\nu'_2}  \rho(t') \hc U(t,t') P_{\nu_1\nu_2}  }\braket{\ti B_{\nu_1\nu_2}(t) \ti B_{\nu'_1\nu'_2}(t')} \\
+\bT{U(t,t') \rho(t') P_{\nu'_1\nu'_2} \hc U(t,t') P_{\nu_1\nu_2} - P_{\nu_1\nu_2} U(t,t') \rho(t') P_{\nu'_1\nu'_2} \hc U(t,t')  } \braket{\ti B_{\nu'_1\nu'_2}(t')\ti B_{\nu_1\nu_2}(t) }  \Big\},
}
\end{widetext}
and the time-local one becomes
\begin{widetext}
\ml{\label{eq:scatt_timelocal}
S_\mrm{TL} (t) =
-\hb^{-2} \int_{t_0}^t dt' \sum_{\nu_1\nu_2\nu'_1\nu'_2} \Big\{ \\
\bT{ P_{\nu_1\nu_2} U(t,t') P_{\nu'_1\nu'_2} \hc U(t,t') \rho(t)  - U(t,t') P_{\nu'_1\nu'_2}  \hc U(t,t') \rho(t)  P_{\nu_1\nu_2}  }\braket{\ti B_{\nu_1\nu_2}(t) \ti B_{\nu'_1\nu'_2}(t')} \\
+\bT{ \rho(t) U(t,t')  P_{\nu'_1\nu'_2} \hc U(t,t') P_{\nu_1\nu_2} - P_{\nu_1\nu_2}  \rho(t) U(t,t') P_{\nu'_1\nu'_2} \hc U(t,t')  } \braket{\ti B_{\nu'_1\nu'_2}(t')\ti B_{\nu_1\nu_2}(t) }  \Big\}.
}
\end{widetext}




\section{The Hamiltonian}\label{app:fundamental_Hamil}
In this appendix we describe the steps needed to obtain the Hamiltonian used in the main text, starting from a more fundamental Hamiltonian. The fundamental Hamiltonian is given by
\al{\label{eq:fundamental_Hamil}
H = H_\mrm s + H_\mrm{0,ph} + H_\mrm{e-ph}+H_{\gamma}+H_\kappa + H_\Gamma.
}
The part governing the QD-cavity system is
\ml{
  H_\mrm s = \hb\om_\mrm g \hc c_\mrm g c_\mrm g + \hb\om_\mrm e \hc c_\mrm e c_\mrm e + \hb\om_\mrm{cav} \hc a a \\
+\hb g (\hc a \hc c_\mrm g c_\mrm e + \hc c_\mrm e c_\mrm g a).
}
The free phonon Hamiltonian is
\al{
   H_\mrm{0,ph} = \sum_{\bs k} \hb\om_\mrm{\bs k}\hc b_\mrm{\bs k}b_\mrm{\bs k}.
}
The interaction between the electrons and the phonons is
\al{
H_\mrm{e-ph} = \sum_{\bs k} \bP{M_\mrm{gg}^\mrm{\bs k}\hc c_\mrm g c_\mrm g+M_\mrm{ee}^\mrm{\bs k}\hc c_\mrm e c_\mrm e}(\hc b_\mrm{-\bs k}+b_\mrm{\bs k}).
}
The last three contributions $H_{\gamma}$, $H_\kappa$, and $H_\Gamma$ refer to different reservoirs and their interaction with the system, giving rise to various forms of Markovian decay, which are introduced in the main text. Their explicit forms are not needed and will therefore not be discussed further in this appendix. For an elaboration on the above Hamiltonians, see the main text, \pref{sec:cQED_system}.

We only consider a single electron in the system, hence the following relation holds
\al{\label{eq:one_electron}
\hc c_\mrm g c_\mrm g+\hc c_\mrm e c_\mrm e = 1,
}
which may be used to eliminate the ground state operator $\hc c_\mrm g c_\mrm g$ from the Hamiltonian. Physically, this elimination can be motivated by the fact that we only have one electron in two levels, implying a perfect correlation between the two electronic states and hence it is sufficient to treat one of the levels explicitly. For reasons to be elaborated below, we choose the excited state.

The elimination results in the following changes
\al{
  H_\mrm s = \hb\om_\mrm{eg} \hc c_\mrm e c_\mrm e + \hb\om_\mrm{cav} \hc a a +\hb g (\hc a \hc c_\mrm g c_\mrm e + \hc c_\mrm e c_\mrm g a),
}
where a constant energy term has been removed and we introduced the transition frequency of the QD, defined as $\om_\mrm{eg} = \om_\mrm{e}-\om_\mrm{g}$. Furthermore, we get
\ml{\label{eq:Heph_tmp1}
H_\mrm{e-ph} = \sum_{\bs k} M^\mrm{\bs k}(\hc b_\mrm{-\bs k}+b_\mrm{\bs k})\hc c_\mrm e c_\mrm e+\sum_{\bs k}M_\mrm{gg}^\mrm{\bs k}(\hc b_\mrm{-\bs k}+b_\mrm{\bs k}),
}
where we have introduced an effective interaction matrix element as $M^\mrm{\bs k}=M_\mrm{ee}^\mrm{\bs k}-M_\mrm{gg}^\mrm{\bs k}$ and the last term without any electron operators corresponds to the phonon interaction with the fully occupied ground state. In thermal equilibrium, before any excitation of the system, the QD is in its ground state and the phonon system is in an equilibrium state that takes into account the presence of the electron in the ground state. We wish to describe a situation that deviates from this thermal equilibrium and therefore it would be advantageous to take into account the phonon interaction with the fully occupied ground state from from very beginning. This may be achieved by shifting the phonon operators \cite{Grodecka2007} through a unitary transformation, defined as
\al{
b_{\bs k} \rightarrow \e{S} b_{\bs k} \e{-S}, \quad S = \sum_{\bs k} \frac{M_\mrm{gg}^\mrm{\bs k}}{\hbar \om_{\bs k}}(\hc b_\mrm{-\bs k}-b_\mrm{\bs k}).
}
This transformation only affects the phonon operators and leads to the substitution in the total Hamiltonian
\al{
b_{\bs k} \rightarrow b_{\bs k} - \frac{M_\mrm{gg}^\mrm{-\bs k}}{\hbar \om_{\bs k}}.
}
This removes the last term in \pref{eq:Heph_tmp1} and introduces a new term given by $-\sum_{\bs k} \bS{2M^\mrm{\bs k}M_\mrm{gg}^\mrm{-\bs k}}/\bS{\hbar \om_{\bs k}}\hc c_\mrm e c_\mrm e$, which yields a simple energy renormalization that can be absorbed into the bare excited state energy $\hbar \om_\mrm e$.

For describing single photon emission, it is sufficient to operate in a one-excitation subspace of the QD-cavity Hilbert space. As a specific basis, we choose the following: $\bT{\ket 1 = \ket{\mrm e, n=0},\ket 2 = \ket{\mrm g, n=1},\ket 3 = \ket{\mrm g, n=0}}$. Along with the reformulations introduced above, projecting the second quantized Hamiltonian on to this subspace changes the following parts of the total Hamiltonian defined in \pref{eq:fundamental_Hamil}
\al{\label{eq:Hs_tmp1}
  &H_\mrm s = \hb\om_\mrm{eg}\sig_{11}+ \hb\om_\mrm{cav}\sig_{22} +\hb g (\sig_{12}+\sig_{21}),\\
  &H_\mrm{e-ph} = B\sig_{11},
}
where $B=\sum_{\bs k} M^\mrm{\bs k}(\hc b_\mrm{-\bs k}+b_\mrm{\bs k})$ and $\sig_{pq} = \ket p \bra q$.

To simplify the equations, the QD-cavity detuning $\Delta = \om_\mrm{eg} - \om_\mrm{cav}$ can be introduced into \pref{eq:Hs_tmp1}. This can be achieved by moving into a rotating frame described by the unitary operator $T(t)=\exp (-i\om_\mrm{cav}[\sig_{11} +\sig_{22}]t)$, through which we obtain the Hamiltonian
\al{\label{eq:Hs_tmp2}
  H_\mrm s = \hb\Delta\sig_{11} +\hb g (\sig_{12}+\sig_{21}).
}

\section{The polaron transformed Hamiltonian}\label{app:polaron_trans_hamil}
In this appendix we will apply the polaron transformation to the total Hamiltonian.

The total Hamiltonian presented in \pref{app:fundamental_Hamil} contains contributions from reservoirs needed to include the Markovian losses. For the final equations, the decay terms arising from these Hamiltonians will, however, not be affected by the polaron transformation introduced in this appendix and they will be omitted in the following. We explicitly demonstrate this in \pref{app:POL_lindblad}. The Hamiltonian is
\ml{\label{eq:Hs_untranny}
H = \hb\Delta\sig_{11} +\hb g (\sig_{12}+\sig_{21}) \\
+ \sig_{11}\sum_{\bs k} M^\mrm{\bs k}(\hc b_\mrm{-\bs k}+b_\mrm{\bs k}) + \sum_{\bs k} \hb\om_\mrm{\bs k}\hc b_\mrm{\bs k}b_\mrm{\bs k}.
}
The transformation we apply is known as the polaron transformation \cite{Wurger1998,Wilson-Rae2002,Brandes2005,Hohenester2010} and is defined in the following way
\al{
\bar O = \e{S} O \e{-S}
}
where
\al{
S &= \sig_{11}C,\\
C &= \sum_{\bs k} \lambda_{\bs k}(\hc b_\mrm{-\bs k}-b_\mrm{\bs k}),\quad \lambda_{\bs{k}} = \frac{M^\mrm{\bs k}}{\hbar \om_{\bs k}}.
}

For performing the transformation we employ the Baker-Campbell-Hausdorff formula which states
\ml{
\bar O = \e S O \e{-S} \\
= O + \bS{S,O} + \frac{1}{2!}\bS{S,\bS{S,O}} + \frac{1}{3!}\bS{S,\bS{S,\bS{S,O}}} + \cdots
}
The transformed operators are:
\al{
\bar \sig_{11} = \sig_{11}, \quad \bar \sig_{12} = \sig_{12}\e C,\quad \bar b_{\bs k} = b_{\bs k} -\lambda_{-\bs k}\sig_{11}.
}
Inserting these expressions and simplifying the resulting Hamiltonian yields
\al{\label{eq:H_tranny1}
\bar H = \hb\Delta\sig_{11} +\hb g (\sig_{12}X_++\sig_{21}X_-) +\sum_{\bs k} \hb\om_\mrm{\bs k}\hc b_\mrm{\bs k}b_\mrm{\bs k},
}
where the detuning has been redefined as
\al{\label{eq:polaron_shift}
\Delta \rightarrow \Delta-\sum_{\bs k}\abs{M^\mrm{\bs k}}^2/(\hbar^2 \om_{\bs k})
}
to take into account the so-called polaron shift of the $\ket 1$ state and further we introduced the phonon operators
\al{\label{eq:Xpm_def}
X_\pm=\e{\pm C}.
}

While \pref{eq:H_tranny1} is still an exact representation of the original Hamiltonian, the electron-photon and electron-phonon interactions have now been mixed into a single term. One might say that the photons now interact with a polaron, the electron-phonon quasi-particle, instead of the bare electron. It would be advantageous to more clearly separate the electron-photon and the electron-phonon interaction. To achieve this
separation\cite{Wurger1998,Wilson-Rae2002} we replace $X_\pm$ with $X_\pm+\braket X-\braket X$ in \pref{eq:H_tranny1} to obtain
\al{\label{eq:H_tranny3}
\bar H = \bar H_{\mrm s'} + \bar H_{\mrm s'-\mrm{ph}'} + H_{\mrm{0,ph}},
}
with
\begin{subequations}
\al{
\bar H_{\mrm s'} &= \hb\Delta\sig_{11}+\hb g \braket X(\sig_{12}+\sig_{21}), \\
\bar H_{\mrm s'-\mrm{ph}'} &= \hb g (\sig_{12}\delta X_++\sig_{21}\delta X_-), \\
H_{\mrm{0,ph}} &= \sum_{\bs k} \hb\om_\mrm{\bs k}\hc b_\mrm{\bs k}b_\mrm{\bs k}.
}
\end{subequations}
where $\braket X$ is defined in \pref{eq:X_avg_def} and $\delta X_\pm$ in \pref{eq:deltaXpm_def}. Now $\bar H_{\mrm s'}$ contains what might be referred to as a system Hamiltonian, however, it is not the original system consisting of only the electron and photon, as the phonon quantity $\braket X$ has entered. It is, however, of great advantage to include this term in the new system Hamiltonian, since then photon processes are treated to all order as well as preserving the detailed balance condition \cite{Carmichael1973}. This would not be case if the system Hamiltonian were defined as the first term in \pref{eq:H_tranny1}, thereby ending up treating the photons only to second order \cite{Hohenester2009c,Hohenester2010}. The quantity $\braket X$ has the effect of renormalizing the light-matter coupling strength $g$. From its definition, \pref{eq:X_avg_def}, it is clear that $0<\braket X \leq 1$, and hence the presence of the phonons will always decrease the effective light-matter coupling. The Hamiltonian $\bar H_{\mrm s'-\mrm r'}$ contains the interaction between the system and reservoir, which has been made weaker by the introduction of the difference operators $\delta X_\pm$, making it more suitable for a treatment using perturbation theory.

\section{Lindblad decay rates under the polaron transformation}\label{app:POL_lindblad}
In this appendix we will calculate the effect of the polaron transformation on a typical Lindblad decay rate. We consider the radiative contribution to the background QD decay rate, which has complicated non-radiative contributions as well, which can not be treated in a simple manner. Our starting point is the Hamiltonian
\ml{
H=\hb\om_\mrm{eg}\sig_\mrm{ee} + \sum_l\hb\Omega_l\hc a_la_l + \sum_{\bs k} \hb\om_\mrm{\bs k}\hc b_\mrm{\bs k}b_\mrm{\bs k}\\
+\sig_\mrm{ee}\sum_{\bs k} M^\mrm{\bs k}(\hc b_\mrm{-\bs k}+b_\mrm{\bs k})+
\sum_l \hb g_l (\hc a_l \sig_\mrm{ge}+a_l \sig_\mrm{eg}),
}
describing a two-level QD with ground and excited states, $\bT{\ket g, \ket e}$, coupled to a phonon bath given by the $b_\mrm{\bs k}$ operators and a photon bath given by the $a_l$ operators. Applying the polaron transformation as described in \pref{app:polaron_trans_hamil}, we obtain
\ml{
H'=\hb\om'_\mrm{eg}\sig_\mrm{ee} + \sum_l\hb\Omega_l\hc a_la_l + \sum_{\bs k} \hb\om_\mrm{\bs k}\hc b_\mrm{\bs k}b_\mrm{\bs k}\\
+\sum_l \hb g_l (\hc a_l X_+\sig_\mrm{ge}+a_l X_- \sig_\mrm{eg})
}
where $\om'_\mrm{eg}$ includes the polaron shift and $X_\pm$ is defined in \pref{eq:Xpm_def}. We now divide the transformed Hamiltonian as follows
\al{
H'=H'_0+H'_\mrm I,
}
where the free part is
\al{
H'_0=\hb\om'_\mrm{eg}\sig_\mrm{ee} + \sum_l\hb\Omega_l\hc a_la_l + \sum_{\bs k} \hb\om_\mrm{\bs k}\hc b_\mrm{\bs k}b_\mrm{\bs k},
}
and the interaction part is
\al{
H'_\mrm I&=\sum_l \hb g_l (\hc a_l X_+\sig_\mrm{ge}+a_l X_- \sig_\mrm{eg})\\
&=B\sig_\mrm{ge}+\hc B\sig_\mrm{eg},
}
where we have defined the combined photon-phonon operator $B$ as
\al{
B=\sum_l \hb g_l \hc a_l X_+=AX_+.
}
In the original frame the initial condition is assumed to be a fully factorized state
\al{
\chi(0)=\rho_\mrm{QD}(0)\otimes R_\mrm{phonon} \otimes R_\mrm{photon},
}
where $\chi(t)$ is the density matrix of the total system. Performing the polaron transformation on the initial density matrix entangles the QD and phonon operators, so that the initial state no longer remains fully factorized. This complicates the further application of the Reduced Density Matrix formalism and is often neglected under the assumption that it is small \cite{Brandes2005}. Employing this approximation we proceed with the following density matrix in the polaron frame
\al{
\chi'(0)\approx\rho_\mrm{QD}(0)\otimes R_\mrm{phonon} \otimes R_\mrm{photon}.
}
We now follow the standard procedure and can write down the EOM for the excited state population of the QD using \pref{eq:scatt_timelocal}
\al{
\pd t n(t) = -\hb^{-2}\int_0^tdt'\bS{\e{i\om'_\mrm{eg}(t-t')}\braket{\ti B(t-t')\hc B}+\mrm{c.c.}}n(t).
}
From the assumption of a factorized density matrix we obtain
\al{
\e{i\om_\mrm{eg}(t-t')}\braket{\ti B(t-t')\hc B}=
\braket{\ti{X}_+(t-t')X_-}G(t-t'),
}
where the polaron correlation function $\braket{\ti{X}_+(t-t')X_-}$ is given in \pref{eq:XaXb_tmp1} and the photon correlation function is
\al{
G(t-t')=\sum_l [\hb g_l]^2\e{-i(\Omega_l-\om'_\mrm{eg})(t-t')}.
}
If $g_l$ is approximately constant near $\Omega_l=\om'_\mrm{eg}$ one has
\al{
G(t-t')=\hb^2\Gamma \delta(t-t'),
}
where $\Gamma$ is the photon-induced decay rate of the QD, while we neglect the photon Lamb shift. The equation for the QD decay now becomes
\al{
\pd t n(t) = -\Gamma\int_0^tdt'\bS{\delta(t-t')\braket{\ti X_+(t-t')X_-}+\mrm{c.c.}}n(t),
}
where, due to the appearance of the delta function in the integrand we may use for the phonon correlation function
\al{
\braket{\ti X_+(t-t')X_-}\lvert_{t=t'}=\braket{\ti X_+(0)X_-}=1,
}
Therefor, within the stated approximations the polaron transformation does not influence Lindblad decay rates.

\section{Properties of the phonon operators}\label{app:prop_phon_op}
In this appendix we give various results related to the phonon operator arising from the polaron transformation
\al{
X_\pm=\e{\pm C},\quad C = \sum_{\bs k} \lambda_{\bs k}(\hc b_\mrm{-\bs k}-b_\mrm{\bs k}),\quad \lambda_{\bs{k}} = \frac{M^\mrm{\bs k}}{\hbar \om_{\bs k}}.
}
The operators $X_\pm$ may be written in terms of so-called displacement operators \cite{Walls2008}
\al{
D_{\bs k}(\alpha) = \exp\bP{\alpha\hc b_{\bs k}-\alpha^*b_{\bs k}}.
}
If we rewrite the operator $C$ in the following way
\al{
C = \sum_{\bs k} \lambda_{\bs k}(\hc b_\mrm{-\bs k}-b_\mrm{\bs k}) = \sum_{\bs k} (\lambda_{-\bs k}\hc b_\mrm{\bs k}-\lambda^*_{-\bs k}b_\mrm{\bs k}),
}
we can write
\al{\label{eq:Xpm_tmp1}
X_\pm = \prod_{\bs k}\exp\bS{\pm(\lambda_{-\bs k}\hc b_\mrm{\bs k}-\lambda^*_{-\bs k}b_\mrm{\bs k})} = \prod_{\bs k} D_{\bs k}(\pm\lambda_{-\bs k}).
}
We will need the following useful properties \cite{Walls2008,Glauber1963} of the displacement operators
\al{\label{eq:dis_rel_1}
\hc D_{\bs k}(\alpha) &= D_{\bs k}^{-1}(\alpha)=D_{\bs k}(-\alpha),\\\label{eq:dis_rel_2}
D_{\bs k}(\alpha)D_{\bs k}(\beta) &= D_{\bs k}(\alpha+\beta)\exp \bP{i\mrm{Im}\bS{\alpha\beta^*}},\\\label{eq:dis_rel_3}
\braket{D_{\bs k}(\alpha)} &= \exp \bP{-\abs{\alpha}^2\bS{n_{\bs k}+1/2}}.
}
In the last expression
\al{\label{eq:avg_phon_nk}
n_{\bsk}=\braket{\hc b_\mrm{\bs k}b_\mrm{\bs k}}=\frac{1}{\ee \bP{\beta \hb \omm{\bsk}}-1}
}
is the average thermal occupation of phonons in mode $\bsk$ and $\beta=(k_\mrm B T)^{-1}$ is the inverse thermal energy. The brackets $\braket{\cdots}=\Tr_\mrm{ph}\bT{\rho_{\mrm{ph},0} \cdots}$ denote the expectation value with respect to the thermal density operator for the phonons
\al{\label{eq:rho_ph_0}
\rho_{\mrm{ph},0} &= \frac{\exp (-\beta H_{\mrm{ph},0})}{\Tr_\mrm{ph} \bT{\exp (-\beta H_{\mrm{ph},0})}},
}
which can be written as a product of the density matrices for the individual $\bs k$ modes as $ \rho_{\mrm{ph},0} = \prod_{\bs k}\rho^{\bs k}_{\mrm{ph},0}$, where
\al{\label{eq:rho_ph_k_0}
\rho^{\bs k}_{\mrm{ph},0} =\frac{\exp (-\beta \hb\om_{\bs k}\hc b_{\bs k}b_{\bs k})}{\Tr_\mrm{ph,\bsk} \bT{\exp (-\beta \hb\om_{\bs k}\hc b_{\bs k}b_{\bs k})}},
}
is the density matrix for the $\bs k$th phonon mode.

The first property we will derive is
\al{
\braket{X_\pm(t)} = \braket{X_\pm} = \braket{X},
}
where the time-evolution is with respect to $H_{\mrm{ph},0}$, resulting in the standard expression for free evolution
\al{\label{eq:bk_time_evol}
b_{\bs k}(t)=\e{-i\om_{\bs k}t}b_{\bs k}.
}
Combining Eqs. (\ref{eq:Xpm_tmp1}) and (\ref{eq:bk_time_evol}) and taking the thermal expectation value using eqs. (\ref{eq:rho_ph_0}) and (\ref{eq:rho_ph_k_0}) we get\cite{phonon_factor_operators}
\al{
\braket{X_\pm(t)} = \prod_{\bs k} \braket{D_{\bs k}(\pm\e{i\om_{\bs k}t}\lambda_{-\bs k})},
}
where the individual terms in the product may be evaluated using \pref{eq:dis_rel_3}, yielding
\al{\notag
\braket{X_\pm(t)} &= \prod_{\bs k} \exp \bP{-\abs{\lambda_{\bs k}}^2\bS{n_{\bs k}+1/2}} \\\notag
&= \exp \bP{-\sum_{\bs k}\abs{\lambda_{\bs k}}^2\bS{n_{\bs k}+1/2}}\\\label{eq:X_avg_def}
&=\braket{X},
}
being independent of time.

Next we will evaluate the polaron correlation functions, defined as
\begin{subequations}\label{eq:Bpm_def}
\al{
B_+(t,t')=\braket{\delta X_\pm(t)\delta X_\pm(t')},\\
B_-(t,t')=\braket{\delta X_\pm(t)\delta X_\mp(t')},
}
\end{subequations}
where
\al{\label{eq:deltaXpm_def}
\delta X_\pm(t) = X_\pm(t)-\braket{X}.
}
Inserting this into the definitions of $B_\pm(t-t')$ we easily find
\al{\notag
\braket{\delta X_a(t)\delta X_b(t')}&=\braket{(X_a(t)-\braket{X})(X_b(t)-\braket{X})}\\
&=\braket{X_a(t)X_b(t')}-\braket{X}^2,
}
indicating that we only need to evaluate $\braket{X_a(t)X_b(t')}$, where $a,b=\pm$. From the above we get
\al{
X_a(t)X_b(t')= \prod_{\bs k}D_{\bs k}(a\e{i\om_{\bs k}t}\lambda_{-\bs k})D_{\bs k}(b\e{i\om_{\bs k}t'}\lambda_{-\bs k}),
}
and using \pref{eq:dis_rel_2} allows us to write
\al{
X_a(t)X_b(t')= \prod_{\bs k}D_{\bs k}(\lambda_{-\bs k}[a\e{i\om_{\bs k}t}+b\e{i\om_{\bs k}t'}])\\
\times\exp \bS{abi\abs{\lambda_{\bs k}}^2\sin (\om_{\bsk} (t-t'))}.
}
Taking the thermal average and employing \pref{eq:dis_rel_3} yields
\ml{\label{eq:tmpeq1}
\braket{X_a(t)X_b(t')}= \exp \bT{-\sum_{\bsk}\abs{\lambda_{\bs k}}^2 (2n_{\bsk}+1)}\\
\times \exp \bT{-ab \varphi(t-t')},
}
where we have defined the function
\begin{widetext}
\al{\label{eq:phi_def}
\varphi(t-t')&=\sum_{\bsk}\abs{\lambda_{\bs k}}^2\bT{[2n_{\bsk}+1]\cos(\om_{\bsk} [t-t'])-i\sin (\om_{\bsk} [t-t'])}\\
&=\sum_{\bsk}\abs{\lambda_{\bs k}}^2\bT{n_{\bsk}\e{i\om_{\bsk} [t-t']}+[n_{\bsk}+1]\e{-i\om_{\bsk} [t-t']} }
}
\end{widetext}
Comparing Eqs. (\ref{eq:tmpeq1}), (\ref{eq:X_avg_def}), and (\ref{eq:phi_def}) we see that
\al{\label{eq:XaXb_tmp1}
\braket{X_a(t)X_b(t')}&= \braket{X}^2 \e{-ab \varphi(t-t')},\\
\braket X &= \e{-\varphi(0)/2}.
}
Going back to \pref{eq:Bpm_def} and using \pref{eq:XaXb_tmp1}, we obtain the final result
\al{\label{eq:Bpm_final_result}
B_\pm(t,t')=B_\pm(t-t')=\braket{X}^2\bP{\e{\mp\varphi(t-t')}-1},
}
where, as expected, the equilibrium phonon correlation functions depend only on the time-difference and not the absolute time. We will also be needing $B_\pm(t',t)$, i.e., with the time arguments interchanged. These functions are available through complex conjugation
\al{\notag
&\braket{\delta X_a(t)\delta X_b(t')}^*=\\\notag
&\bS{\Tr_\mrm{ph}\bT{\rho_{\mrm{ph},0}\delta X_a(t)\delta X_b(t')}}^*=\\\notag
&\Tr_\mrm{ph}\bT{\bS{\rho_{\mrm{ph},0}\delta X_a(t)\delta X_b(t')}^\dagger}=\\\notag
&\Tr_\mrm{ph}\bT{\delta X_{\bar b}(t')\delta X_{\bar a}(t)\rho_{\mrm{ph},0}}=\\\notag
&\braket{\delta X_{\bar b}(t')\delta X_{\bar a}(t)},
}
where the bar signifies multiplication by $-1$ and we used $(X_+)^\dagger=X_-$. Finally, we obtain the following relation
\al{
B_\pm(t',t)=\bS{B_\pm(t,t')}^*.
}

\section{Scattering rates in the polaron frame}\label{app:EOM_polaron}
In this appendix we explicitly define the scattering rates in the polaron frame entering in \pref{eq:POL_scatt_terms}. The building blocks are the integrals $K_{nmkl}^\pm(t)$ defined in \pref{eq:Knmkl_def_main_paper}. Viz the discussion in \pref{sec:degree_of_nonmarkov} we take the long-time limit in the integrals, hence we define
\al{
K_{nmkl}^\pm \equiv K_{nmkl}^\pm(\infty).
}
The rates are
\al{
\Gamma_1 &= 2\mrm{Re}\bS{K_{2211}^-+K_{2112}^+},\\
\Gamma_2 &= 2\mrm{Re}\bS{K_{1122}^-+K_{1221}^+},\\
\gamma_1 &= K_{1122}^-+\bS{K_{2211}^-}^*+K_{1221}^++\bS{K_{2112}^+}^*,\\
\gamma_2 &= \bS{K_{1122}^-}^*+\bS{K_{1221}^+}^*,\\
\gamma_3 &= \bS{K_{2211}^-}^*+\bS{K_{2112}^+}^*,\\
iG_1 &= K_{2122}^--\bS{K_{2221}^-}^*+K_{2221}^+-\bS{K_{2122}^+}^*,\\
iG_2 &= K_{1112}^--\bS{K_{2221}^-}^*+K_{1211}^+-\bS{K_{2122}^+}^*,\\
iG_3 &= K_{1221}^-+\bS{K_{2112}^-}^*+K_{1122}^++\bS{K_{2211}^+}^*,\\
iG_4 &= -\bS{K_{2221}^-}^*-\bS{K_{2122}^+}^*,\\
iG_5 &= -\bS{K_{1112}^-}^*-\bS{K_{1211}^+}^*.
}




\section{Analytical expression for QD decay rates}\label{app:anal_rates}
In this appendix we derive the analytical expressions for the QD decay rates discussed in \pref{sec:anal_approx}. We proceed in two steps; Firstly, an expression is derived that is valid whenever the cavity can be adiabatically eliminated, and secondly, we take the large detuning limit, which simplifies the phonon induced rates to the expression presented in the main text. We only perform the explicit derivation for the QD decay rate in the original frame, \pref{eq:purcell_orig}, but the derivation for the same quantity in the polaron frame, \pref{eq:purcell_pol}, follows a similar procedure.

From \ref{sec:theory_orig_frame} we get the EOM for the excited QD population
\al{\label{eq:sig11_adia}
\pd t \asig{11} = -\Gamma \asig{11}+2g\imm\bS{\asig{12}},
}
and the photon-assisted polarization
\ml{
\pd t \asig{12} = -\bS{-i\Delta+\ti\gamma_{12}}\asig{12}\\
-i\bS{g+\mathcal{G}^>}\asig{11}+i\bS{g+\mathcal{G}^<}\asig{22},
}
with $\ti\gamma_{12}=\gamma+\ree[\gamma_{12}]+(\kappa+\Gamma)/2$ and where the long-time limit has been taken in all phonon-induced rates, wherefor we omit the time argument. For the cavity to be adiabatically eliminated it can not perform any back-action on the QD, hence it can not enter in the above EOM for the photon-assisted polarization and we put the cavity population, $\asig{22}$, equal to zero. This is valid in the regime where the cavity decay rate, $\kappa$, is much larger than all other parameters. Furthermore, when the total dephasing time $1/\abs{\ti\gamma_{12}}$ is much shorter than the characteristic timescale for $\asig{11}$, we may put $\pd t \asig{12} = 0$.  From this we get
\al{
\asig{12}=-i\frac{g+\mathcal{G}^>}{-i\Delta+\ti\gamma_{12}}\asig{11},
}
which when inserted in \pref{eq:sig11_adia} yields
\al{
\pd t \asig{11} =-\Gamma_\mrm{tot}\asig{11}
}
where the total QD decay rate is defined as
\al{
\Gamma_\mrm{tot} = \Gamma +2g\frac{\ti\gamma_{12}}{\ti\gamma_{12}^2+\Delta^2}\bS{g+\ree[\mathcal{G}^>]-\frac{\Delta}{\ti\gamma_{12}}\imm[\mathcal{G}^>]}.
}
We are interested in the spontaneous emission rate from the QD due to the coupling to the cavity field, hence we expect the final result to scale with $g^2$. For this reason we expand the QD-cavity evolution operator $U(t)$ up to first order in $g/\Delta$
\ml{
U(t)=\e{-i\Delta t}\sigg{11}+\sigg{22}\\+\frac{g}{\Delta}(\e{-i\Delta t}-1)(\sigg{12}+\sigg{21})+\mathcal{O}((g/\Delta)^2).
}
Using this expansion and \pref{eq:Ggl_def} we find
\al{
\ree[\mathcal{G}^>]\propto \frac{\imm[D^>(\om=0)-D^>(\om=\Delta)]}{\Delta},
}
which is small compared to the remaining terms and will be neglected. From the expansion of the time evolution operator we also find that $\ree[\gamma_{12}]$ scales as $g^2$, which makes it a higher order effect that can be neglected. We finally arrive at \pref{eq:purcell_orig}
\al{
\Gamma_\mrm{tot} = \Gamma +2g^2\frac{\gamma_\mrm{tot}}{\gamma_\mrm{tot}^2+\Delta^2}\bS{1+\frac{\hbar^{-2}}{\gamma_\mrm{tot}}\ree[D^>(\om=\Delta)]},
}
where $\gamma_\mrm{tot}=\gamma+(\kappa+\Gamma)/2$ is the total dephasing rate.




\begin{thebibliography}{63}
\bibitem [{QD_()}]{QD_phonon_parameters}%
  \BibitemOpen
  \href@noop {} {}\bibinfo {note} {We take typical GaAs parameters:
  $d=5370~\mrm{kg}\mrm{m}^{-3}$, $c_\mrm s=5110~\mrm{m}\mrm{s}^{-1}$, $D_\mrm e
  = -14.6~\mrm{eV}$, $D_\mrm g = -4.8~\mrm{eV}$,
  $l_{xy,\mrm{e}}=6.18~\mrm{nm}$, $l_{xy,\mrm{g}}=4.48~\mrm{nm}$, and
  $l_\mrm{eff,z}=8~\mrm{nm}$. Due to the hard wall confinement in the $z$
  direction the length $l_\mrm{eff,z}$ should be considered as an effective
  length.}\BibitemShut {Stop}%
\bibitem [{pol()}]{pol_lindblad_rates}%
  \BibitemOpen
  \href@noop {} {}\bibinfo {note} {While performing the polaron transformation
  we omit the contributions to the Hamiltonian giving rise to the Lindblad
  decay terms. Even though these Hamiltonians are not invariant under the
  polaron transformation, the fact that the corresponding reservoirs are
  assumed to be memory-less or delta correlated in time, has the consequence
  that the polaron transformation has no effect on the resulting decay terms.
  Also see \pref{app:polaron_trans_hamil}.}\BibitemShut {Stop}%
\bibitem [{pho()}]{phonon_factor_operators}%
  \BibitemOpen
  \href@noop {} {}\bibinfo {note} {As both $X_\pm(t)$ and $\rho_{\mrm{ph},0}$
  factor into individual $\bs k$ contributions, the averaging may be performed
  for one $\bs k$ at a time and no crossterms in e.g. $\bs k$ and $\bs k'$
  occur.}\BibitemShut {Stop}%
\end{thebibliography}


%



\end{document}